\documentclass[aps,prstab,twocolumn,groupedaddress,nobibnotes,nofootinbib,showpacs,titlepage]{revtex4}
\usepackage{epic,eepic}
\usepackage{epsfig}
\usepackage{pspicture}
\usepackage{graphicx}
\usepackage{amscd,xy}           
\usepackage{subfigure}
\usepackage{floatflt}           
\usepackage{rotating}
\usepackage{picins}
\usepackage[varumlaut]{yfonts}
\usepackage{array}              
\usepackage{amsmath,amssymb,amsthm,amsfonts,bm} 
\usepackage{mathpazo}
\usepackage{revsymb}
\usepackage{draftcopy}
\usepackage{color}
\usepackage{indentfirst}  
\usepackage{pslatex}
\usepackage{latexsym}
\usepackage{textcomp}     
\usepackage{longtable}    %
\usepackage{dcolumn}      
\usepackage{xspace}
\usepackage{setspace}
\usepackage{paralist}
\usepackage{url}
\usepackage{exscale,relsize}
\usepackage[hang,bf]{caption}
\include{rgb}
\newenvironment{rcase}{\left.\begin{aligned}}{\end{aligned}\right\rbrace}
\newenvironment{comp}{\begin{compactenum}[(1)]\vspace{0.2cm}}{\end{compactenum}}
\newcommand\textsfbf[1]{\textsf{\bfseries#1}}
\newcommand{\bef}{\begin{figure}}
\newcommand{\eef}{\end{figure}}
\newcommand{\beq}{\begin{equation}}
\newcommand{\eeq}{\end{equation}}
\newcommand{\beqa}{\begin{equation}\begin{aligned}}
\newcommand{\eeqa}{\end{aligned}\end{equation}}
\newcommand{\beqar}{\begin{equation}\begin{aligned}\begin{rcase}}
\newcommand{\eeqar}{\end{rcase}\end{aligned}\end{equation}}
\newcommand{\bit}{\begin{itemize}}
\newcommand{\eit}{\end{itemize}}
\newcommand{\bnote}{\begin{center}\large\red}
\newcommand{\enote}{\black\normalsize\end{center}}
\newcommand{\bcomp}{\begin{comp}}
\newcommand{\ecomp}{\end{comp}}
\newcommand{\sups}[1]{\raisebox{1ex}{\scriptsize #1}}
\newcommand{\ie}{i.e.,\xspace}
\newcommand{\eg}{e.g.,\xspace}

\newcommand{\macros}{macroparticles~}
\def\Cplusplus{{\rm C\raise.5ex\hbox{\small ++}}}
\DeclareMathAlphabet{\mathpzc}{OT1}{pzc}{m}{it} 
\def\@makefntext#1{%
   \def\baselinestretch{1}%
   \reset@font\footnotesize
   \parindent 1em%
   \noindent
   \hb@xt@1.8em{%
     \Hy@raisedlink{\hyper@anchorstart{footnote@\the\c@footnote}\hyper@anchorend}%
     \hss\@makefnmark
   }%
   #1\par
}%
\graphicspath{{/root/TEX/TH/UNI/PUB/PRSTAB/EPS3/}
              {/root/TEX/LOGOS/}}
\begin{document}
\printfigures
\printtables
\title{Impact of Electric Current Fluctuations
       Arising from Power Supplies and Space Charge
       on Charged-Particle Beams:\\
\textit{A Measurement-Based Stochastic Noise Model of Fermilab's Booster Synchrotron}}
%
\author{Phil S. Yoon}\email{syoon@fnal.gov}
\affiliation{\vspace{0.1in}
 Fermi National Accelerator Laboratory, Batavia, IL 60510, U.S.A.\\
 Department of Physics and Astronomy, 
 University of Rochester, Rochester, NY 14620, U.S.A.}
\date{\today}
\begin{abstract}
Electric current fluctuations are 
one type of unavoidable machine imperfections 
and induce magnetic-field perturbation
as a source of instabilities in accelerators.
This paper presents the measurement-based modeling 
methodology of fluctuating electric current 
arising from the power system of 
Fermilab's Booster synchrotron 
to discuss the ramifications of 
the presence of ripple current and
space-charge defocusing effects.
We also present the method of generating stochastic noise
and the measurement and analysis methods 
of ripple current and offending electromagnetic 
interferences residing in the Booster power system.
This stochastic noise model, accompanied by 
a suite of beam diagnostic calculations, 
manifests that the fluctuating power-supply current, 
when coupled to space charge and impinging upon a beam,
can substantially enhance beam degradation phenomena---such as
emittance growth and halo formation---during the Booster 
injection period.
With idealized and uniform charge-density distribution,
the fractional growth of rms emittances 
due to ripple current under space charge 
turn out to be about $8~\sim~9~\%$ 
in both transverse planes 
over the injection period of 2.2 ms
prior to beam acceleration.
\end{abstract}
\maketitle
\newpage
%
\section{\label{sec:intro}INTRODUCTION AND MOTIVATION}
\begin{figure}[h!]
  \vspace{0.2cm}
  \centering
    \includegraphics[scale=0.40]{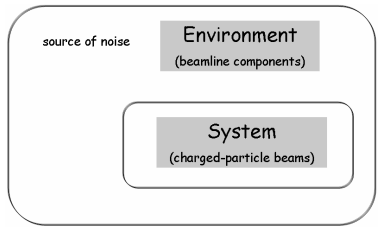}
    \caption{\label{fig:syst-surr}System and surroundings}
\end{figure}

%
As is common in other natural systems, 
subtle fluctuations are ubiquitous and inevitable 
in particle accelerator systems.
An ensemble of charged particles is defined as a {\em system}, 
and all the beamline components 
(magnets, power supplies, RF cavities, beam position monitors, etc.)
for accelerating, guiding, and diagnosing particle beams 
as {\em environment}, or {\em surroundings}. 
The system of a charged-particle beam perceives 
the environment of beamline components 
as a source of noise as illustrated by 
FIG.~\ref{fig:syst-surr}.
External noise is intrinsic to particle accelerators of all types 
due to unavoidable machine imperfections; 
\eg ripple current from power supplies, ground vibration motion, etc.
%
%
After being motivated by earlier findings from 
an idealized and simplified theoretical model~\cite{haloamp:clb},
we speculated that the adverse influence of 
power-supply current fluctuations possibly account for 
beam loss phenomena observed during the injection process 
of the Booster.    
Hence, an independent and novel approach 
has been made to build a \textit{realistic} 
stochastic noise model, based upon a series of power-supply noise\footnote{
~In the present context, we will use the terms \textit{(power-supply)~noise} 
and \textit{current fluctuations} and \textit{ripple current} 
interchangeably.} 
measurements, 
to investigate and explore the impact of current fluctuations 
on charged-particle beams of the Booster at injection energy of 400 MeV.
%
\section{\label{sec:methodology}MODELING METHODOLOGY}
%
%
For investigations of the impact of power-supply 
current fluctuations on charged-particle beams 
in an accelerator lattice structure,
we began with building a preliminary noise model.
As a first step, we designed and added 
a new physics module for generating 
stochastic noise to 
the existing ORBIT-FNAL package~\cite{orbit:pac1999}.
%
The new noise module is capable of generating 
a wide spectrum of stochastic noise
employing the Ornstein-Uhlenebck stochastic process~\cite{o-u}
that is governed by 
a Langevin-like stochastic differential equation~\cite{vankampen}.
%
\par 
Prior to detailed experimental measurements, 
we corroborated with the preliminary noise model
using a linear lattice
that non-white, or colored noise could possibly 
enhance beam degradation process of our interest.
These preliminary findings are consistent with 
earlier findings from the theoretical model 
of collective space-charge modes coupled with 
dynamic noise~\cite{haloamp:clb}.
%
As a next step, in addition to adding the new noise module 
to the ORBIT-FNAL, the existing {\em TeaPot module} 
was upgraded to establish Fermilab's Booster ring 
using a TEAPOT-style~\cite{teapot}~Booster lattice, 
and the{\em~Diagnostic module} 
with new parallelized beam diagnostic calculations: 
\textit{actions, halo amplitudes, etc.}
\begin{figure*}[ht!]
   \centering
   \includegraphics[scale=1.7]{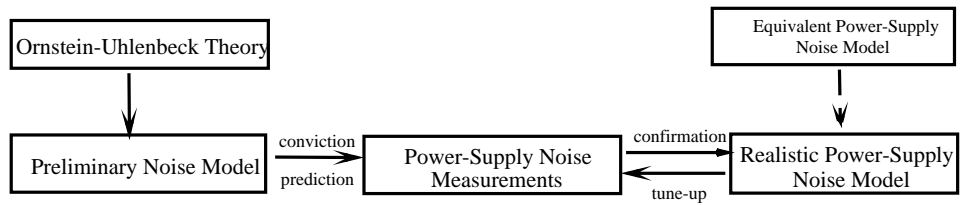}
   \caption{\label{fig:model-procedure}
            Multi-stage approach to modeling 
            the impact of power-supply noise 
            on a beam with realism}
\end{figure*}
\begin{figure}[ht!]
 \begin{center}
    \includegraphics[scale=0.50]{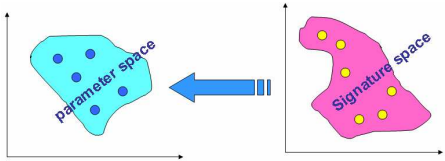}
   \caption{\label{fig:signal-param}
    Mapping from an experimental signature space 
    to a stochastic parameter space}
 \end{center}
\end{figure}
%
\par As confirmed with the preliminary model, 
we proceeded to devise methods for 
direct measurements of 
common-mode and differential-mode voltages, 
and ripples in the electric current. 
Repeated measurements and Fourier analysis 
confirm that a substantial amount of noise, 
which can be transmitted to the magnet system,
is indeed present in the power system.
%
Moreover, we performed equivalent-circuit simulations 
to investigate any offending resonances that can 
float around the magnet system. 
%
Based upon the measurement data and 
the results of Fourier analysis, 
stochastic parameterization of Booster 
ripple current is performed by means of 
matching power spectral densities 
between measured ripple currents and modeled 
Ornstein-Uhlenbeck (O-U) noise~\cite{o-u}.
%
While translating modeled O-U noise to 
induced magnetic-field fluctuations, 
we tracked macroparticles 
in the presence of 3-D space-charge effects.
%
\par The following FIG.~\ref{fig:model-procedure}~and~FIG.~\ref{fig:signal-param}
illustrate the multi-stage approach to the stochastic noise modeling.
For the purpose of the stochastic parameterization of 
ripple current, mapping from an experimental~\textit{signature space} 
to a stochastic~\textit{parameter space} was done;
the preliminary model was faithfully tuned up 
with the power-supply noise measurements.
As a consequence, we managed to match 
FFT power spectral densities 
between physical noise and modeled O-U noise.
%
\section{\label{sec:classification}CLASSIFICATIONS OF NOISE}
%
In general, noise can be categorized into two types:
\textit{external noise} and \textit{internal noise}.
In our stochastic noise model, 
ripple current arising from 
Gradient Magnet Power Supply (GMPS) units 
of the Booster are regarded to be 
fluctuating external influences 
acting on the Booster beam
(see FIG.~\ref{fig:syst-surr}).
%
\subsection{\label{subsec:external}External Noise}
External noise is originated from a source outside of the system;
that is, beamline components. 
Since the effects of external noise can be described by 
a stochastic differential equation (SDE)\footnote{
~A stochastic differential equation is 
a differential equation (DE) in which 
the coefficients are random functions of 
the independent variable, which is usually \textit{time}.},
we chose Langevin Equation (LE) as SDE for 
indeterministic current fluctuations 
arising from each GMPS unit\footnote{
The Booster power supply will be frequently referred to as GMPS.}.
It should be noticed that power-supply ripple currents 
are considered \textit{indeterministic}, or~\textit{random}, 
or~\textit{aperiodic}, in the sense that 
it never exactly repeats itself.
	\subsection{\label{subsec:emi}Electromagnetic Interferences}  
Electromagnetic-Interference (EMI) noise~\cite{mohan,horowitz} 
results from rapid changes in voltage and current 
in a power supply. 
Transmissions of EMI noise are characterized as 
either {\em radiative}, or {\em conductive}.
Conductive EMI noise, such as differential-mode (DM) and
common-mode (CM) noise, is usually several orders of 
magnitude higher than the radiative EMI, 
and can be more harmful to the system.
Given impedance ($Z(\omega)$) as a function of frequency~$\omega$,
fluctuations in common-mode voltage ($V_{CM}$) create 
common-mode current ($I_{CM}$), 
in addition to the inherent ripple current arising from 
sudden potential changes in the power-supply system.
The EMI problem is thereby worsened and could result in
larger current fluctuations, or common-mode current,
or severe system damage.
%
\section{\label{sec:process}STOCHASTIC PROCESS}
%
Of several different stochastic processes\footnote{
In this paper, we use terms {\em stochastic process}, {\em stochastic function},
{\em random process}, {\em random function} 
and {\em process} interchangeably.}, 
such as Poisson process, Wiener process, etc.,
we chose the Ornstein-Uhlenbeck process 
to represent electrical current fluctuations
as classified in the preceding subsection.
%
\subsection{\label{subsec:langevin}Langevin Equation}
    In 1908, after the formulation of the Brownian movement by
    Einstein and Smoluchowski~\cite{einstein,einstein:brownian}, 
    P.~Langevin introduced the concept of 
    the equation of motion of a stochastic variable 
    (\ie the position coordinate of a Brownian particle)~\cite{langevin}.
    Langevin Equation (LE) is considered to be 
    the first example of a Stochastic Differential Equation (SDE)\footnote{
    A stochastic differential equation is 
    a differential equation with a stochastic (random) term.
    Therefore, its solution is also a random function.}.
    Langevin wrote the equation of motion for a Brownian particle 
    according to Newton's second law under the assumption that 
    a Brownian particle is subject to two forces:
    \textit{damping force ($\mathcal{F}_{d}$)} and 
    \textit{fluctuating force ($\mathcal{F}_{f}$)} 
    \begin{equation}\label{eqn:total-force}
      \mathbb{F}(t)~=~\mathcal{F}_{d}(t) + \mathcal{F}_{f}(t)
    \end{equation}
    \begin{equation}\label{eqn:langevin-1}
      \mathpzc{m}~\frac{d\sups{2}\mathpzc{x}(t)}{dt\sups{2}}~ 
      =~\underbrace{-~\zeta~\frac{d\mathpzc{x}(t)}{dt}}_{\text{systematic force:~dissipation}}
      ~+~\underbrace{\mathcal{F}_{f}(t)}_{\text{stochastic force:~fluctuation}}
    \end{equation}
    Here, $\mathbb{F}(t)$, $\mathpzc{m}$, $\mathpzc{x}$, and $\zeta$ represent 
    the total force, particle mass, displacement, and the friction coefficient, respectively.
    The first term ($-\zeta~\dot{\mathstrut\mathpzc{x}}(t)$\footnote{
    ~The notations, $\dot{\mathpzc{x}}$ and $\ddot{\mathpzc{x}}$, denote 
    $d\mathpzc{x}/dt$ and $d^{2}\mathpzc{x}/dt^{2}$, respectively.}) 
    on the right-hand side of Eqn.~(\ref{eqn:langevin-1}) represents 
    the viscous drag as a function of time, or dynamic friction. 
    The second term $\mathcal{F}_{f}(t)$ represents 
    fluctuations which could be from white noise\footnote{ 
    ~{\em White noise} is noise with a flat frequency spectrum.}
    or non-white noise.  
    The form of Eqn.~(\ref{eqn:langevin-1}) can be transformed to 
    Eqn.~(\ref{eqn:langevin-2}) of first order.
    For modeling physical noise of the Booster power system, 
    we employed {\em non-white} noise, or {\em off-white} noise
    in our investigations.
   \begin{equation}\label{eqn:langevin-2}
     \dot{\mathstrut v}(t) + \alpha v(t) = \mathcal{L}(t),
   \end{equation}
    where $\mathcal{L}(t)$ is a stochastic driving force,
    and $\alpha$ represents $\zeta$/$m$.
    The following assumptions are made about 
    the fluctuation part $\mathcal{L}(t)$:
\begin{compactenum}[(1)]\itemsep 0.2cm
    \item $\mathcal{L}(t)$ is a function of time only,
          and independent of $\mathpzc{x}$. 
    \item zero-mean distribution;
          ~\begin{equation}\label{eqn:zero-mean}\langle\mathcal{L}(t)\rangle = 0 \end{equation}
    \item The variation rate of $\mathcal{L}(t)$ 
          is much faster than the velocity 
          of a Brownian particle, $v(t)$.
          Hence, the autocorrelation function $\mathcal{C}_{\mathcal{L}}(t,~t^{\prime})$
          is of Dirac-$\delta$ function;~\ie 
\end{compactenum} 
\begin{equation}\label{eqn:acf-white} 
   \langle\mathcal{L}(t)\mathcal{L}(t^{\prime})\rangle 
 = \mathcal{A}\delta(t-t^{\prime})
\end{equation}
\vspace{0.2cm}
    The expressions above define the statistical properties 
    of $\mathcal{L}(t)$. There is a great advantage in 
    using LE instead of using 
    Fokker-Planck Equation (FPE) of mathematical complex.
    The Langevin's method is much easier 
    to understand than the FPE since
    it is based upon the time evolution of 
    a stochastic variable, whereas the FPE applies 
    to the time evolution of the probability distribution. 
    As such, LE allows us to dispense with
    the calculation of the diffusion coefficient, 
    thus reducing associated mathematical complications. 
    As a consequence, we built an effective 
    but far more simplified model of stochastic noise.
%
\subsection{\label{subsec:o-u}Ornstein-Uhlenbeck Process}
%
As in Eqn.~(\ref{eqn:langevin-2}), 
LE for Brownian motion is given as,
\begin{equation}\label{eqn:langevin-3}
  \dot{\mathstrut \xi}(t) + \omega\xi(t) = \mathcal{L}(t)
\end{equation}
As explained in the preceding subsection,
LE is associated with $\delta$-correlated 
Gaussian stochastic forces
of statistical properties.
This stochastic process $\xi(t)$ is referred to as an O-U process.
Note that the noise strength $\mathcal{A}$ 
does not dependent upon the variables $\xi$.
Since the O-U stochastic process
is inherently to represent the velocity of a Brownian particle,
it is the appropriate choice of stochastic process 
for modeling electric noise, or current fluctuations~\cite{o-u}; 
\ie a time-derivative of electric charge ($dQ/dt$).
Both position ($x(t)$) and velocity ($v(t)$) describe 
Langevin's Brownian motion. 
However, by utilizing the O-U process of 
the velocity of a Brownian particle,
LE can be reduced to a $1^{st}$-order 
linear stochastic differential equation
that is derived from Newton's $2^{nd}$ law.
As a result, 
LE as a $1^{st}$-order SDE
is straightforward to find solutions.\\
%
%
The O-U process is associated with an exponentially-decreasing 
autocorrelation function $\mathcal{C}_{\xi}(t,~t^{\prime})$~\cite{taylor} 
and a finite autocorrelation time~$\tau_{ac}$~\cite{o-u}\footnote{
The autocorrelation function~$\mathcal{C}(t,~t^{\prime})$
determines the property of a stochastic process of interest.}.
\begin{equation}\begin{aligned}\label{eqn:o-u-acf}
  \mathcal{C}_{\xi}(t,~t^{\prime})
  &= \langle~\xi(t)\,\xi(t^{\prime})~\rangle 
  = \mathcal{A}\exp(-\omega_{ac}\,\vert t~-~t^{\prime}\vert),\\
  &\text{where~$\tau_{ac} = \omega_{ac}^{-1}$.}
\end{aligned}\end{equation}   
in which $\xi(t)$, $\omega_{ac}$, and $\mathcal{A}$ are 
a stochastic function, an autocorrelation frequency, 
and a constant noise strength, respectively.
O-U processes are associated with 
the following probability density function:
\begin{equation}\label{eqn:o-u-pdf}
  \mathcal{W}_{1}(\xi,~t) 
   =\frac{1}{\sqrt{\pi\mathcal{A}/\omega_{ac}}}\cdot 
   \exp\Bigl(-\frac{1}{2}\frac{\xi\sups{2}}{\mathcal{A}/2\omega_{ac}}\Bigr)
\end{equation}
According to the Doob's theorem~\cite{doob}, 
the O-U process is the only stochastic process 
with all of the following properties:
 (1) stationary process, 
 (2) Gaussian   process,
 (3) Markovian  process.
In particular, if a process is invariant to 
translations in time
(e.g. a shift in time ($\alpha$)) then 
the process is called a 
\textit{stationary process}~\cite{wang,vankampen}.
For a stationary process, 
we can make the following simplifications:
     \begin{equation}\label{eqn:stationary}
  	 \begin{split}
     		\langle~\xi(t_{1}~+~\alpha)\xi(t_{2}+\alpha)
                ~\dotsb~\xi(t_{n}~+~\alpha)~\rangle 
              = \langle~\xi(t_{1})\xi(t_{2})
                ~\dotsb~\xi(t_{n})~\rangle
         \end{split}
     \end{equation}
   where $\xi(t)$ is the stochastic function and $\langle~\ldots~\rangle$ 
   is the  statistical average.
   The form of Eqn.~(\ref{eqn:stationary}) implies the followings:
 \begin{compactenum}[(1)]
   \vspace{-0.2cm}
   \item Since the underlying mechanisms causing the fluctuations 
         do not change with time, the stochastic properties of 
         a stationary process are conserved. 
   \item The important parameter in the O-U process 
         is \textit{relative time} and not the absolute time.
         \begin{equation}\label{eqn:stationary-3}
             \langle~\xi(t_{1}-\alpha)\xi(t_{1})~\rangle 
              = \langle~\xi(t_{1})\xi(t_{1}~+~\alpha)~\rangle
         \end{equation}
         Therefore, the autocorrelation function 
         $\mathcal{C}_{\xi}(t,~t^{\prime})$ for a stationary process 
         is a function of $\vert t - t^{\prime}\vert$ only. 
   \item The ensemble average and the time average are 
         the same, which leads to the ergodic property.
 \end{compactenum}
%
\subsection{\label{subsec:markov}Markov Process}
Markov processes involve the use of {\em conditional probability}.
%
   \begin{equation}\label{eqn:cond-prob}
     \mathcal{W}_{2}(\xi_{1},~\xi_{2},~t) 
   = \mathcal{W}_{1}(\xi_{1})\mathcal{P}_{2}(\xi_{1}~\vert~\xi_{2},~t)
   \end{equation}
  The Markov process is therefore defined as follows~\cite{vankampen}:
  \begin{equation}\label{eqn:markov-3}
    \mathcal{P}_{n}\bigl(~\xi_{1}~t_{1},~\xi_{2}~t_{2},~\ldots,
    ~\xi_{n-1}~t_{n-1}~\vert~\xi_{n}~t_{n}~\bigr)
     = \mathcal{P}_{2}(~\xi_{n-1}~t_{n-1}~\vert~\xi_{n}~t_{n}~)   
  \end{equation}
  The form of Eqn.~(\ref{eqn:markov-3}) implies that 
  all the $\mathcal{P}_{n}$ for $n > 2$ can be derived, 
  when only $\mathcal{P}_{2}$ is known.
  In other words,
  \begin{center}
     {\em Only the present condition determines the future condition.}
  \end{center}
In order to avoid unnecessary mathematical complexity 
in building a stochastic noise model, 
we exploited the Markovian property. 
This is another reason why, of all the stochastic processes,
the O-U process is the most convenient choice for
modeling the Booster ripple current. 
%
\section{\label{sec:non-white-noise}NON-WHITE NOISE GENERATION}
%
%
\subsection{\label{subsec:stochastic-noise}Stochastic Properties}
%
  Langevin Equation governs an O-U process.  
  Hence, if we use an O-U process
  to model GMPS current fluctuations,
  we need to solve LE. 
  By solving the convenient $1^{st}$-order linear LE,
  we were able to extract more statistical properties 
  of the O-U process besides~
  Eqns.~(\ref{eqn:zero-mean}),~(\ref{eqn:acf-white}),
  ~and~(\ref{eqn:o-u-pdf}).
   \par Let us first consider a $1^{st}$-order SDE
   of the form of LE.
   \begin{equation}\label{eqn:langevin-3}
      \dot{\mathstrut \xi}(t) = f(\xi) + \eta(t)
   \end{equation}
Here $\eta(t)$ is non-white Gaussian noise 
with the autocorrelation function 
$\mathcal{C}_{\eta}$:
  \begin{equation}\label{eqn:acf-couleur}
    \mathcal{C}_{\eta}(t,~t^{\prime}) 
  = \langle \eta(t)\eta(t^{\prime}) \rangle
  = \frac{\mathcal{A}}{2\omega_{ac}}\exp(-\omega_{ac}\vert t - t^{\prime}\vert)
  \end{equation}
Non-white noise $\eta$ is governed by LE 
with a white-noise driving force of $\mathcal{L}(t)$:
   \begin{equation}\label{eqn:langevin-4}
      \dot{\mathstrut \eta}(t) + \omega_{ac}\eta(t) = \mathcal{L}(t)
   \end{equation}
The autocorrelation function $\mathcal{C}_{\mathcal{L}}$ is
$\delta$-correlated with a strength $\mathcal{A}$:
  \begin{equation}\label{eqn:acf-blanc}
     \mathcal{C}_{\mathcal{L}}(t,~t^{\prime}) 
   = \langle \mathcal{L}(t)\mathcal{L}(t^{\prime}) \rangle
   = \mathcal{A}\delta(t - t^{\prime})                    
  \end{equation}
   Ornstein and Uhlenbeck~\cite{o-u}, 
   Doob~\cite{doob}, and van Kampen~\cite{vankampen} 
   use the integration method to find the statistical 
   properties of non-white noise, or colored noise
   from LE.
   We, on the other hand, solve LE
   as a 1$^{st}$-order DE.
   The general solution of a 1$^{st}$-order 
   inhomogeneous DE is a linear superposition of 
   a homogeneous solution ($\eta_{h}$) 
   and a particular solution ($\eta_{p}$). 
Hence,
   \begin{equation}\label{eqn:sde-solution}
      \begin{split}
         \eta(t) &= \eta_{h}(t) + \eta_{p}(t) \\
                 &= \eta(0)\cdot\exp(-\omega_{ac}~t) + 
	            \int_{0}^{t}\,ds\cdot\exp(-\omega_{ac}(t-s))\cdot\mathcal{L}(s)
      \end{split}
   \end{equation}
   From Eqn.~(\ref{eqn:sde-solution}), 
   the stochastic process at the next time step 
   \begin{math}~t~+~\Delta t~\end{math} can be obtained.
   \begin{equation}\label{eqn:next-step-1}
      \begin{split}
       &\eta(t~+~\Delta t)\\ 
       &= \eta(0)\exp(-\omega_{ac}(t~+~\Delta t))\\ 
       &+ \int_{0}^{t~+~\Delta t}ds~
         \exp(-\omega_{ac}(t~+~\Delta t - s))\mathcal{L}(s) \\            
       &= \exp(-\omega_{ac}\Delta t) \eta(t) 
         + \underbrace{\int_{t}^{t~+~\Delta t}ds~
          \exp(-\omega_{ac}(t~+~\Delta t~-~s))
          \mathcal{L}(s)}_{\text{$\mathcal{H}(t,~t~+~\Delta t)$}}
     \end{split}
   \end{equation}
   Let $\mathcal{H}(t, t~+~\Delta t)$ 
   be the second term of Eqn.~(\ref{eqn:next-step-1}).
   \begin{equation}\label{eqn:2.10}
      \mathcal{H}(t,~t~+~\Delta t)
      ~\equiv~\int_{t}^{t~+~\Delta t}\,ds \cdot 
      \exp(-\omega_{ac}(t~+~\Delta t~-~s))
      \cdot \mathcal{L}(s)    
   \end{equation}
    By transforming the variables of integration,
   we can obtain
   \begin{equation}\label{eqn:h}
     \begin{split}
      \mathcal{H}(0,~\Delta t) 
      &= \int_{0}^{\Delta t} \,d\tilde{s} 
         \cdot \exp(-\omega_{ac}(\Delta t 
         - \tilde{s}))\cdot \mathcal{L}(\tilde{s}~+~t) \\ 
      &= \exp(-\omega_{ac}\Delta t) 
         \cdot \int_{0}^{\Delta t}\,d\tilde{s}\cdot 
         \exp(\omega_{ac}\tilde{s})
         \cdot \mathcal{L}(\tilde{s}~+~t)
     \end{split}      
   \end{equation}
By squaring~Eqn.~(\ref{eqn:h}), we arrive at
\begin{widetext}
   \begin{equation}\label{eqn:h2}
     \begin{split}
      &\mathcal{H}^{2}(0,~\Delta t)\\ 
      &= \exp(-2\omega_{ac}\Delta t) 
         \int_{0}^{\Delta t}\int_{0}^{\Delta t}d\tilde{s}\,d\tilde{s}\sups{$\prime$}
         \exp(\omega_{ac}(\tilde{s}~+~\tilde{s}\sups{$\prime$})) 
         \mathcal{L}(\tilde{s}~+~t)
         \mathcal{L}(\tilde{s}\sups{$\prime$}~+~t)
     \end{split}
  \end{equation}
\end{widetext}
  The statistical properties of a random variable
  can be investigated by the calculations of various moments.
  We calculate the first and the second central moments\footnote{
  ~When a mean value of a variable is included 
  in the moment calculation, 
  it is referred to as \textit{central moment}.}
  by averaging Eqns.~(\ref{eqn:h}) and~(\ref{eqn:h2})  
  over an ensemble of particles.
  The first two moments determine 
  the complete statistical properties of 
  the O-U noise because it is a zero-mean Gaussian process. 
  For zero-mean Gaussian,
  the $1\sups{st}$ moment vanishes.
  \begin{equation}\label{eqn:ave-h}
      \langle~\mathcal{H}(0,\Delta t)~\rangle = 0
  \end{equation}
  Accordingly, keeping in mind that the O-U process 
  is a stationary process, 
  the $2\sups{nd}$ moments boil down to  
  \begin{equation}\label{eqn:2nd-moments}
     \begin{split}
       &\langle~\mathcal{H}^{2}(0,\Delta t)~\rangle\\
       &= \exp(-2\omega_{ac}\Delta t)
          \int_{0}^{\Delta t}\,\int_{0}^{\Delta t}d\tilde{s}\,d\tilde{s}\sups{$\prime$}
          \exp(\omega_{ac}(\tilde{s}~+~\tilde{s}\sups{$\prime$}))\langle~\mathcal{L}(\tilde{s})
          \mathcal{L}(\tilde{s}\sups{$\prime$})~\rangle\\
       &= \mathcal{A}\exp(-2\omega_{ac}\Delta t)
          \int_{0}^{\Delta t}\,d\tilde{s}\exp( 
          2\omega_{ac}\tilde{s})\\
       &= \frac{\mathcal{A}}{2\omega_{ac}}\lbrace 1 
         - \exp( -2\omega_{ac}\Delta t)\rbrace
     \end{split} 
  \end{equation}
The second moments of $\mathcal{H}$
can be expanded in a closed form 
as in Eqn.~(\ref{eqn:2nd-moments-expand}).
\begin{equation}\begin{aligned}\label{eqn:2nd-moments-expand}
  &\langle~\mathcal{H}^{2}(0,~\Delta t)~\rangle\\
  &= \frac{\mathcal{A}}{2\omega_{ac}}\Bigl\lbrace~1~-~\exp(-2\omega_{ac}\Delta t)\Bigr\rbrace\\ 
  &= \frac{\mathcal{A}}{2\omega_{ac}}
     \Bigl\lbrack~2(\omega_{ac}\Delta t)~-~2(\omega_{ac}\Delta t)^{2} 
   + \frac{8}{3!}(\omega_{ac}\Delta t)^{3} 
   - \ldots~\Bigr\rbrack \\
  &= \mathcal{A}\Delta t
     \Bigl\lbrack~1~-~\mathcal{R}_{t} + \frac{2}{3}\mathcal{R}_{t}^{2} 
   - \frac{1}{3}\mathcal{R}_{t}^{3}   +~\ldots~\Bigr\rbrack,
\end{aligned}\end{equation}
with $\mathcal{R}_{t}$ being $\omega_{ac}\Delta t$.
What determines the $\langle~\mathcal{H}^{2}~\rangle$
is $\omega_{ac}\Delta t$, which is time step $\Delta t$
in units of autocorrelation time~$\tau_{ac}$, not
autocorrelation time, or time step by itself.
   This module is designed to generate 
   O-U stochastic noise $\eta(t)$
   that is to be applied to macroparticles 
   in the form of magnetic-field perturbation:
   \textit{~autocorrelation time 
   ($\tau_{ac}$), time step ($\Delta t$), and
   noise strength ($\mathcal{A}$)}.
%
\subsection{\label{subsec:box-muller}Box-Muller-Like Transformation}
\begin{figure*}[htb!]
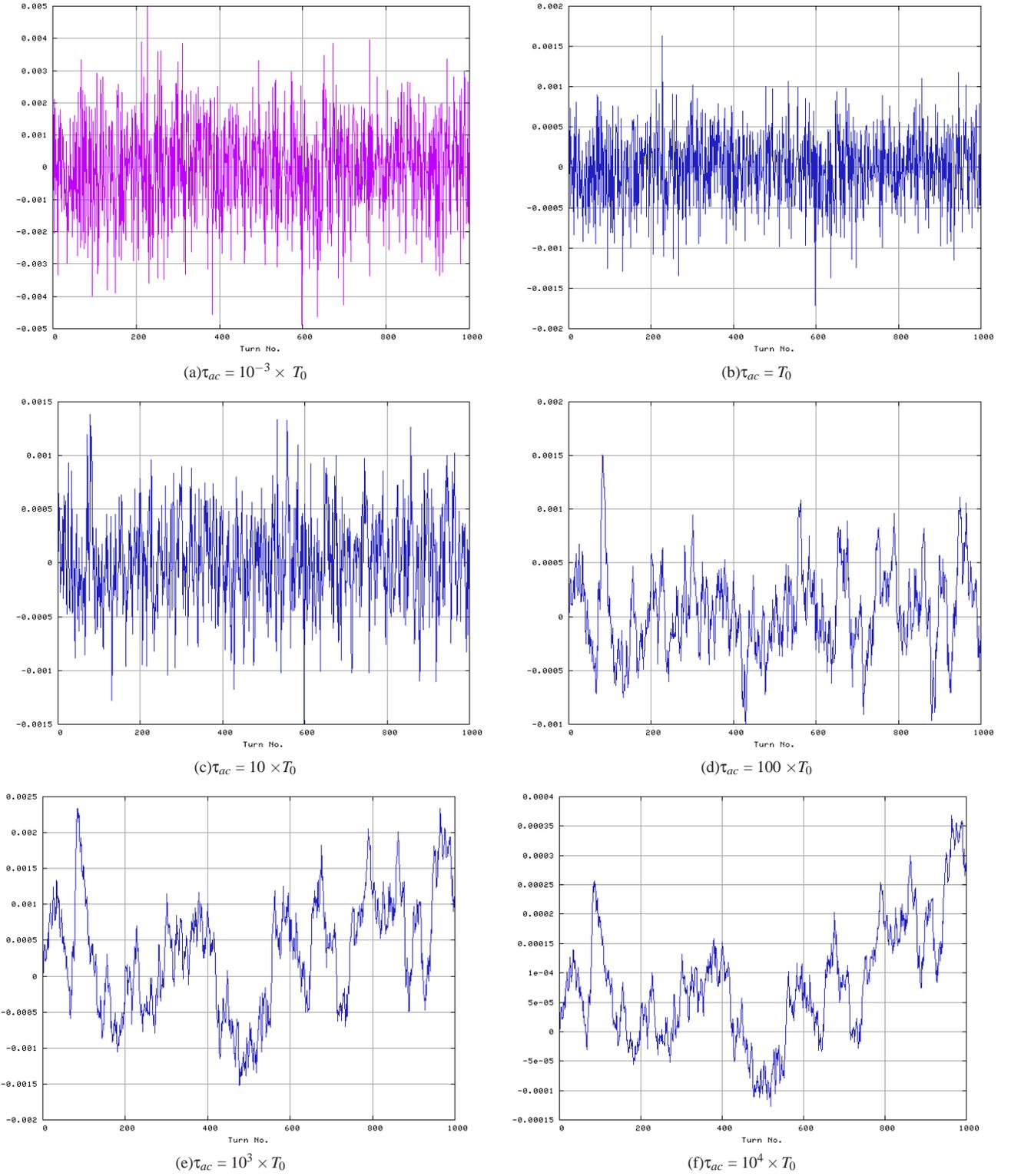

  \begin{center}
     \subfigure[$\tau_{ac}$ = $10^{-3}\times~T_{0}$]
     {\includegraphics[scale=0.33]{samplePath-1em3T0.eps3}}\hspace{0.2in}
     \subfigure[$\tau_{ac}$ = $T_{0}$]
     {\includegraphics[scale=0.33]{samplePath-T0.eps3}}\hspace{0.2in}
     \subfigure[$\tau_{ac}$ = 10 $\times T_{0}$]
     {\includegraphics[scale=0.33]{samplePath-10T0.eps3}}\hspace{0.2in}
     \subfigure[$\tau_{ac}$ = 100 $\times T_{0}$]
     {\includegraphics[scale=0.33]{samplePath-100T0.eps3}}\hspace{0.2in}
     \subfigure[$\tau_{ac}$ = $10^{3} \times T_{0}$]
     {\includegraphics[scale=0.33]{samplePath-1kT0.eps3}}\hspace{0.2in}
     \subfigure[$\tau_{ac}$ = $10^{4} \times T_{0}$]
     {\includegraphics[scale=0.33]{samplePath-10kT0.eps3}}\hspace{0.2in}
     \caption{\label{fig:sample-paths}
              Sample paths of the Ornstein-Uhlenbeck noise 
              over 1,000 tracking turns; 
              the autocorrelation time ($\tau_{ac}$) 
              ranges from $10^{-3}\times~T_{0}$
              to $10^{4}\times~T_{0}$, where $T_{0}$ denotes one revolution period;
              the horizontal axis is turn number and the vertical axis is noise amplitude.}
   \end{center}
\end{figure*}

The Box-Muller (BM) transformation~\cite{box,recette} is 
intrinsically for generating independent Gaussian white noise
---which is a limiting case of physical noise---
from independent uniform random deviates.
In order to generate exponentially-driven Gaussian stochastic noise,
an exponential factor, $\exp(-\omega\Delta t)$ is first multiplied
by the stochastic noise $\eta(t)$ at present time $t$.~
Then, a root-mean-square (rms) value of $\mathcal{H}(0,~\Delta t)$
is added to compute the noise at the next time step $t~+~\Delta t$. 
  \begin{equation}\begin{aligned}\label{eqn:next-step}
    \eta(t~+~\Delta t) 
  &= \exp(-\omega\Delta t)\cdot\eta(t) 
  + \mathcal{C}_{W}\cdot\sqrt{\langle\mathcal{H}(t,~t~+~\Delta t)^{2}\rangle}\\
  &= \exp(-\omega\Delta t)\cdot\eta(t) 
  + \mathcal{C}_{W}\cdot\sqrt{\langle\mathcal{H}(0,\Delta t)^{2}\rangle},
  \end{aligned}
  \vspace{0.1in}
  \end{equation}
where $\mathcal{C}_{W}$ denotes random deviates 
from a rectangular distribution (or white noise).
What Eqn.~(\ref{eqn:next-step}) implies is that 
to generate $\eta(t~+~\Delta t)$,
one needs to know $\eta(t)$ only.
This takes advantage of the powerful {\em Markov property} 
of the O-U process in numerical calculations.
Upon providing with stochastic parameters, 
the variant of the BM transformation is capable of
generating a wide spectrum of stochastic noise: 
\textit{~colored noise, non-white noise, off-white noise, etc.} 
Sample paths of different noises that are generated from
the new noise module are plotted in FIG.~\ref{fig:sample-paths}: 
the autocorrelation time ($\tau_{ac}$) 
ranges from 10\sups{-3}$\times T_{0}$ 
($T_{0}$ denotes one revolution period.)
to $10\sups{4}\times T_{0}$.
The time step is fixed at one revolution period 
at the Booster injection energy.
FIG.~\ref{fig:sample-paths} demonstrates that 
the autocorrelation time governs the pattern of sample path. 
It is therefore evident that the pattern of all sample paths 
are~\textit{aperiodic}.
More details of the non-white noise algorithm 
can be found elsewhere~\cite{thesis:yoon}.
%
\subsection{\label{subsec:noise-app}Application of Noise to Macroparticles}
Since current fluctuations are directly proportional 
to magnetic-field fluctuations, 
in the noise model the ripple-current measurements
are translated into magnetic-field fluctuations 
as in Eqn.~(\ref{eqn:field-perturbation}).
\begin{equation}\label{eqn:field-perturbation}
   \widetilde{\mathstrut\textbf{K}}_{imag}
 = \textbf{K}_{imag} + \Bigl\vert\Delta\textbf{K}_{imag}\Bigr\vert 
 = \textbf{K}_{imag}\cdot\Bigl(1 + \Bigl\vert\Delta\textbf{K}_{imag}\Bigr\vert/\textbf{K}_{imag}\Bigr), 
\end{equation}
where $imag$ denotes magnet index for 
differentiating between each individual main magnets.
In order to distinguish field fluctuations at each type of magnet 
($F$, or $D$), $\textbf{K}_{imag}$ is factored out, and 
the amount of field variation ($\Delta\textbf{K}_{imag}$) 
is normalized by $\textbf{K}_{imag}$ as a perturbation term.
\cleardoublepage
It should be noted that according to experimental measurements 
(see section~\ref{sec:expt}), the amount of ripple current ($\Delta I$) 
is positive above the baseline of a sinusoidal current waveform.
Hence, absolute values of $\Delta\textbf{K}_{imag}$
($\Bigl\vert\Delta\textbf{K}_{imag}\Bigr\vert$)
are taken to represent measured ripple current 
as in Eqn.~(\ref{eqn:field-perturbation}).
%
\section{\label{sec:gmpss}GRADIENT-MAGNET POWER-SUPPLY (GMPS) SYSTEM}
%
The Gradient Magnet Power Supply (GMPS) System  
for the Booster synchrotron powers 
a total of 96 main gradient magnets\footnote{
~The gradient magnet is referred to as 
the combined-function magnet of the Booster.}.
A resonance system is selected in order to 
reduce the size and the cost of the power-supply system. 
The Booster magnet system consists of 48 LC-resonant magnet cells.
A focusing magnet (F), a defocusing magnet (D), a choke, 
and a capacitor bank constitute an individual magnet cell.
In turn, 48 focusing and 48 defocusing magnets are connected 
in series by common buses.
Since the gradient magnets are powered by 
four independent power supplies (GMPS)
that are symmetrically inserted in the LC-resonant system,
the GMPS voltages to ground ($V_{+G}$ and $V_{-G}$)
can be kept as low as possible.
Each GMPS drives current at the fundamental frequency 
of 15 Hz through a string of 12 magnet cells.
The GMPS system includes dual three-phase 
Silicon Controlled Rectifier (SCR) bridges 
connected in series, and fed by a 12-phase 13.8-kV bus 
with shunt (or stray) capacitors connected to ground.
The components of the Booster GMPS system are summarized 
in Table~\ref{tab:gmpss}. 
\squeezetable
\begin{table}[hbt!]
      \centering
         \caption{\label{tab:gmpss}The Booster Gradient-Magnet System}
   \begin{tabular}{@{}l r r r r @{}}\\ \hline
       \textsfbf{Component}  & \textsfbf{No.} \\ \hline
       LC-resonant magnet cells  & 48 \\
       focusing magnets          & 48 \\
       defocusing magnets        & 48 \\
       chokes                    & 48 \\
       capacitor banks           & 48 \\
       GMPS                      &  4 \\
       gradient magnets $/$ cell &  2 \\
       choke $/$ cell            &  1 \\
       magnet cells $/$ GMPS     & 12 \\ \hline
   \end{tabular}
\end{table}     
%
\section{\label{sec:expt}NOISE-MEASUREMENT TECHNIQUES AND ANALYSIS}
%
%
\subsection{\label{subsec:cmn-dmn}
Common-Mode Noise and Differential-Mode Noise}
\begin{figure}[hbt!]
  \includegraphics[scale=0.8]
  {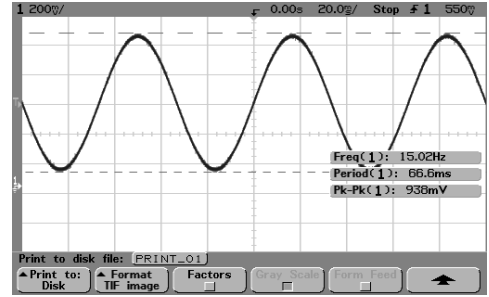}
  \caption{\label{fig:15-hz-current}15-Hz current waveform taken at GMPS \#1}
\end{figure}

\begin{figure}[hbt!]
   \includegraphics[scale=0.1]{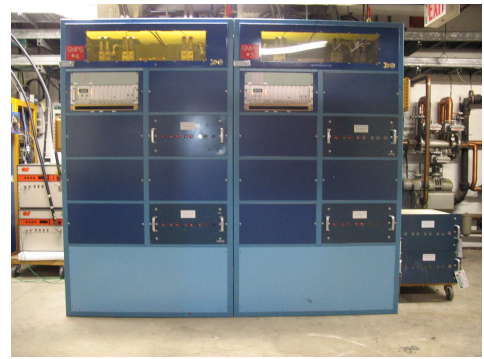}
   \caption{\label{fig:gmps-units}Two GMPS Units}
\end{figure}

\begin{figure}[htb!]
   \begin{center}
     \includegraphics[scale=0.18]{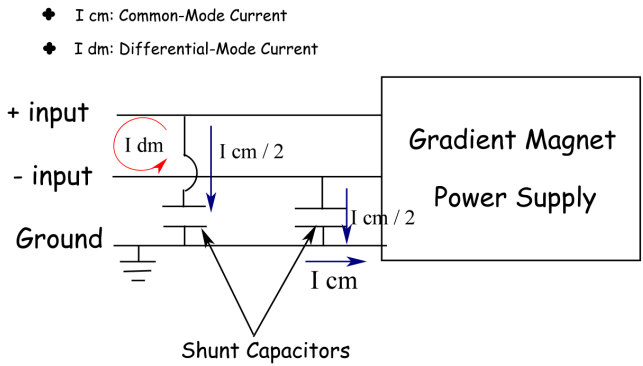}
     \caption{\label{fig:cmn-dmn}
              Common-mode current and differential-mode current 
              at the Booster GMPS}
   \end{center}
 \end{figure}

\begin{figure}[h!]
      \includegraphics[scale=1.0]{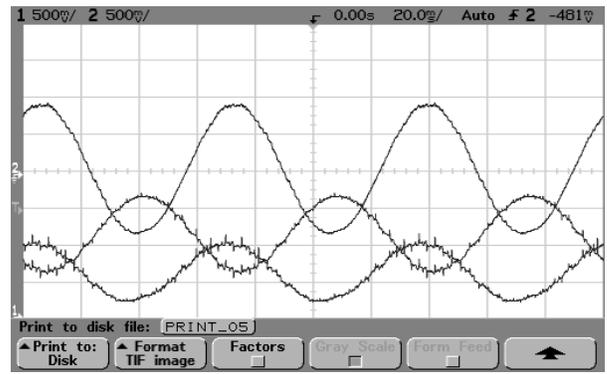}\vspace{0.4in}
      \includegraphics[scale=1.0]{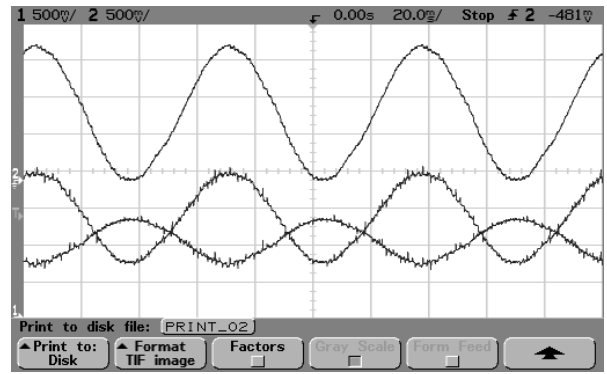}\vspace{0.4in}
      \includegraphics[scale=1.0]{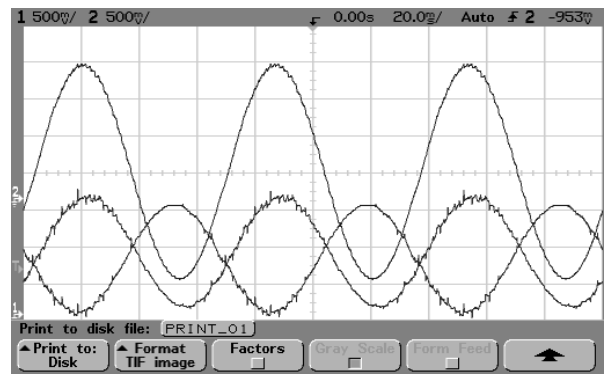}\vspace{0.4in}
      \includegraphics[scale=1.0]{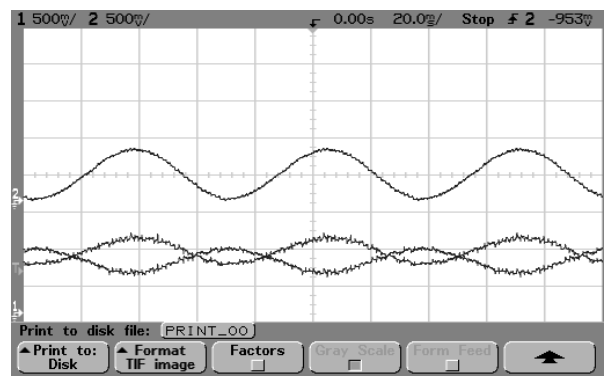}
      \caption{\label{fig:dmv-vpg-vmg}
               The upper waveform is differential-mode voltage ($V_{DM}$)
               that are plotted against a pair of waveforms of $V_{+G}$ and $V_{-G}$.
               Progressing from top to bottom, each plot corresponds to GMPS \#1 through \#4.}
\end{figure}

Starting from summer 2005 through winter 2006, 
we repeatedly conducted direct measurements of 
15-Hz current on the main bus line 
and common-mode and differential-mode voltages 
at each of four GMPS units. 
From a series of measurements, we confirmed that  
ripple current and common-mode voltages 
have consistently been detected and they are not 
of seasonal behavior at each individual GMPS unit.
A 15-Hz current waveform over 3 cycles is shown 
in FIG.~\ref{fig:15-hz-current}.
Frequency and period of the current are shown 
in the shaded boxes on the figure.
Two of four GMPS units located in the East Booster gallery 
are pictured in FIG.~\ref{fig:gmps-units}.
As illustrated by FIG.~\ref{fig:cmn-dmn}, 
the waveforms of both $V_{+G}$ and $V_{-G}$ signals 
are sampled at the two leads on the GMPS control rack. 
Utilizing a digital oscilloscope,\footnote{
~The model name of the digital oscilloscope
used for the measurements is Agilent 54622A, 
and the part number 54622-97014.}
common-mode voltages and differential-mode voltages
are calculated as follows: 
\begin{equation}\begin{aligned}\begin{cases}\label{eqn:vcm-vdm}
  V_{CM} &= V_{+G} + V_{-G}\\
  V_{DM} &= V_{+G} - V_{-G}
\end{cases}\end{aligned}\end{equation}
The waveforms of $V_{+G}$ and 
the inverted $V_{-G}$ ($\widetilde{V}_{-G}$) 
are overlaid for easy comparison on the same scale
in FIG.~\ref{fig:vpg_mvmg}.  
In addition to FIG.~\ref{fig:vpg_mvmg},
overlaid CM voltages are plotted against $V_{+G}$ 
and $\widetilde{V}_{-G}$ for each GMPS
in FIG.~\ref{fig:waveforms-vcm}.
Peak-to-peak measurements quantify 
the potential differences and 
cursor-key-function determines phase lags between two signals. 
Potential differences are displayed 
in FIG.~\ref{fig:pk2pk}.
In principle, the waveforms of $V_{+G}$ and $V_{-G}$ are supposed to be
180 degrees out of phase. However, as displayed in FIG.~\ref{fig:phase-lags}, 
substantial amounts of phase lag are found at each GMPS. 
We calculated the fractional difference 
in amplitudes ($\vert\Delta V/V\vert$) 
by taking the difference 
between $V_{+G}$ and $\widetilde{V}_{-G}$ 
and normalizing the difference by $V_{+G}$.
It was discovered that substantial amounts 
of potential differences in $V_{+G}$
and $V_{-G}$ are present at each GMPS unit. 
The counterparts of DM voltage are shown in FIG.~\ref{fig:dmv-vpg-vmg}.
The DC output of the power supply is filtered with a L-C network,
with the aid of a 15-Hz low-pass filter,
to smooth the differential-mode sawtooth waveform at all GMPS units. 
The waveforms of CM voltage ($V_{CM}$) are noticeably fast fluctuating,
which will induce additional current fluctuations in the system. 
Peak-to-peak amplitudes of $V_{+G}$ and $\widetilde{V}_{-G}$ are plotted
in FIG.~\ref{fig:pk2pk}. Phase lags between waveforms of 
$V_{+G}$ and $\widetilde{V}_{-G}$ are plotted in FIG.~\ref{fig:phase-lags}.
The voltage divisions are set to 500~$mV/div$~and the sweep speed is
set to 20~$mV/div$~in FIG.~\ref{fig:vpg_mvmg} through~FIG.~\ref{fig:pk2pk}. 
From these measurements, it was found that the mismatch of potential differences
and phase lags between $V_{+G}$ and $V_{-G}$ for each GMPS unit are different. 
\par It was found that the following are the two root causes of
common-mode noise arising from each GMPS unit:
 \begin{center}
    \setlength{\fboxsep}{1em}
    \fbox{\begin{minipage}{3.2in}
      \begin{compactenum}[(1)] 
        \item additional phase lags ($\Delta X$) between $V_{+G}$ and $V_{-G}$ 
        \item amplitude (potential) difference between $V_{+G}$~and~$V_{-G}$
      \end{compactenum}
    \end{minipage} }
  \end{center}
When $V_{+G}$ and $V_{-G}$ are added 
in a point-by-point fashion on the scope, 
they do not cancel out each other.
Instead, the ripples on each waveform add up 
and the common-mode voltage thus stands out. 
\par The potential differences and the phase lags measured for 
individual GMPS units are summarized in Table~\ref{tab:voltage-phaselags}. 
Of four GMPS units, fractional potential difference in GMPS \#2 
is the largest and the worst.
\squeezetable
\begin{table}[hbt!]
 \centering
 \caption{\label{tab:voltage-phaselags}
 Difference in voltage amplitudes and phase lags at each GMPS}
   \begin{tabular}{@{}l r r r r @{}}\\ \hline
   \textsfbf{GMPS No.}&\textsfbf{$V_{+G}$} (V)&\textsfbf{$V_{-G}$} (V) & 
       \textsfbf{$\mid\frac{\Delta V}{V}\mid$} &\textsfbf{$\Delta X$}(ms)\\ \hline
       GMPS 1  & 1.577                  & 1.905                 & 20.8 $\%$ & 0.6     \\
       GMPS 2  & 3.232                  & 1.699                 & 47.4 $\%$ & 4.0     \\
       GMPS 3  & 1.598                  & 1.740                 & 8.9  $\%$ & 1.4     \\
       GMPS 4  & 1.581                  & 1.743                 & 10.2 $\%$ & 4.6     \\ \hline
    \end{tabular}
   \end{table}    
FIG.~\ref{fig:current-linear-waveform}~shows 
the ripple current on a linear ramp of the sinusoidal waveform. 
The currents were sampled directly from the magnet bus line. 
Since the ripple currents are such a small fraction of the reference current,
transductor electronics and a current amplifier (TA22 Texas Instruments) 
are used for signal amplification. 
FFT impulses of current waveform of one cycle 
is displayed in FIG.~\ref{fig:fft-current}.
The vertical scale is 20 $dB/div$ and 
the horizontal span is 1,670 Hz.
%
\subsection{\label{subsec:fft}
Power Spectral Density of Noise:~FFT Analysis}
\vspace{-0.2cm}
The measured common-mode voltages from all of four GMPS units 
and the current signal with ripples are Fourier-analyzed. 
To provide \textit{real-time} proof of the presence of 
offending interference in the power-supply system, 
all the signals are analyzed \textit{on the fly} 
without being transported to any commercial software 
for the post-measurement analysis. 
We performed \textit{real-time} analysis 
with the aid of the built-in FFT-function feature 
on the scope. 
As FIG.~\ref{fig:15-hz-current} 
through~FIG.~\ref{fig:current-linear-waveform}, 
the real-time graphics were saved on the scope
at the time of measurements.
The resolution of a resonant peak, or FFT bin size, 
is determined by the FFT sampling rate and 
the number of points.
The number of points on the scope is fixed at 2048,
such that the FFT sampling rates and the span of 
the frequency domain are controlled 
in accordance with the Nyquist sampling theorem. 
In addition, in order to enhance spectrum resolution
around the frequency peak, 
the \textit{Hanning window} was selected 
over flat-top, rectangular, and Hamming windows.
The selected Hanning-window function is given
in Eqn.~(\ref{eqn:hanning}):
\begin{equation}\label{eqn:hanning}
  \mathcal{W}_{H}(t) 
   = \frac{1}{2}
     \Bigl[~1~-~\cos\Bigl(\frac{2\pi\cdot t}{N~-~1}\Bigr)\Bigr],
\end{equation}
in which $\mathcal{W}_{H}(t)$,~$t$~and~$N$ denote 
Hanning-window function, time, and the number of samples, 
respectively. 
\par The following is a list of the FFT settings used on the scope:
\vspace{0.05in}
  \begin{compactenum}[(1)]\itemindent 0.5in \itemsep 0.2cm
    \item FFT sampling rate, $f_{s}$ = 3.53 kSa$/$s
    \item FFT bin size, $\Delta f$ = 1.04 Hz
    \item Frequency-domain span = 1.67 kHz
    \item Horizontal scale = 167 Hz$/$div
    \item Vertical scale = 20 dB$/$div
  \end{compactenum}
\vspace{0.1in}
  \par According to the Nyquist sampling theorem, 
  the oscilloscope determines FFT sampling rate 
  from the chosen span of frequency domain. 
%
\cleardoublepage
 \begin{figure}[h!]
       \includegraphics[scale=1.0]{G1_V+g_-V-g.eps3}\vspace{0.2in}
       \includegraphics[scale=1.0]{G2_V+g_V-g_PRINT_07.eps3}\vspace{0.2in}
       \includegraphics[scale=1.0]{G3_V+g_-V-g_PRINT_18.eps3}\vspace{0.2in}
       \includegraphics[scale=1.0]{G4_V+g_-V-g_PRINT_12.eps3}
       \caption{\label{fig:vpg_mvmg}
       The waveforms of $V_{+G}$ and inverted $V_{-G}$. 
       Progressing from top to bottom,
       each waveform shown on the oscilloscope display 
       corresponds to GMPS \#1 through GMPS \#4.
       As indicated on the upper edge of each display, 
       the voltage division is set to 
       500~mV$\slash$div and sweep speed to 20 ms$\slash$div.}
  \end{figure}
%
  \begin{figure}[h!]
     \begin{center}
       \includegraphics[scale=1.0]{G1_V+g_-V-g_Vcm.eps3}\vspace{0.2in}
       \includegraphics[scale=1.0]{G2_V+g_-V-g_Vcm_PRINT_08.eps3}\vspace{0.2in}
       \includegraphics[scale=1.0]{G3_Vcm_V+g_-V-g_PRINT_19.eps3}\vspace{0.2in}
       \includegraphics[scale=1.0]{G4_Vcm_V+g_-V-g_PRINT_13.eps3}
       \caption{\label{fig:waveforms-vcm}
       The waveforms of $V_{CM}$ are plotted against 
       those of $V_{+G}$ and inverted $V_{-G}$. 
       Starting from top to bottom, each display corresponds to 
       the GMPS \#1 through the GMPS \#4.}
     \end{center}
  \end{figure}
%
  \begin{figure*}[htb!]
     \begin{center}
       \subfigure[\textsfbf{GMPS \#1}]
       {\includegraphics[scale=0.9]{G1_V+g_QuickMeas_PRINT_10.eps3}\hspace{0.2in}
        \includegraphics[scale=0.9]{G1_V-g_QuickMeas_PRINT_11.eps3}}
       \subfigure[\textsfbf{GMPS \#2}]
       {\includegraphics[scale=0.9]{G2_V+g_Quick_PRINT_01.eps3}\hspace{0.2in}
        \includegraphics[scale=0.9]{G2_V-g_Quick_PRINT_03.eps3}}
       \subfigure[\textsfbf{GMPS \#3}]{
       \includegraphics[scale=0.9]{G3_V+g_QuickMeas_15nov2k6.eps3}\hspace{0.2in}
       \includegraphics[scale=0.9]{G3_V-g_QuickMeas_15nov2k6.eps3}} 
       \subfigure[\textsfbf{GMPS \#4}]{
       \includegraphics[scale=0.9]{G4_V+g_QuickMeas_15nov2k6.eps3}\hspace{0.2in}
       \includegraphics[scale=0.9]{G4_V-g_QuickMeas_15nov2k6.eps3}}
       \caption{\label{fig:pk2pk}Peak-to-peak amplitudes (Pk-Pk(1) and Pk-Pk(2)) 
                and frequencies of $V_{+G}$ and inverted $V_{-G}$}
    \end{center}
   \end{figure*}
\cleardoublepage
\begin{figure}[h!] 
       \subfigure[\textsfbf{[GMPS \#1]}: phase lag ($\Delta X$) is 0.6 ms]
       {\includegraphics[scale=0.9]{G1-cursor-06-Dec-2K6.eps3}}
       \subfigure[\textsfbf{[GMPS \#2]} phase lag ($\Delta X$) is 4.0 ms]
       {\includegraphics[scale=0.9]{G2-cursor-06-Dec-2K6.eps3}}
       \subfigure[\textsfbf{[GMPS \#3]} phase lag ($\Delta X$) 1.40 ms]
       {\includegraphics[scale=0.9]{G3_V+g_-V-g_Cursor_print_03-20nov2k6.eps3}}
       \subfigure[\textsfbf{[GMPS \#4]} phase lag ($\Delta X$) 4.60 ms]
       {\includegraphics[scale=0.9]{G4_V+g_-V-g_Cursor_15nov2k6.eps3}}
       \caption{\label{fig:phase-lags} 
        phase lags between $V_{+G}$ and inverted $V_{-G}$}
\end{figure}
%
\cleardoublepage
%
\subsection{\label{subsec:stochastic-param}
Parameterization of GMPS Noise}
\begin{figure}[h!] 
        \includegraphics[width=3.5in,height=2.5in]
        {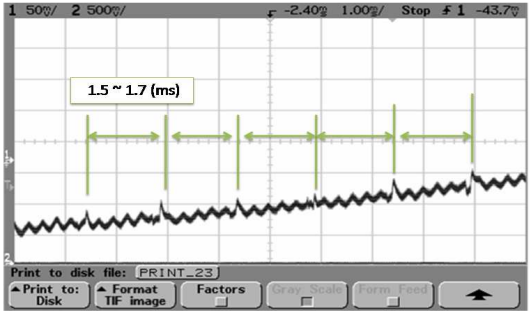}
        \caption{\label{fig:current-linear-waveform}
         ripple current on a linear ramp of the sinusoidal waveform. 
         The current are sampled directly from the magnet bus line.}
  \end{figure}

\begin{figure}[hbt!]
    \begin{center}
    \includegraphics[scale=0.23]{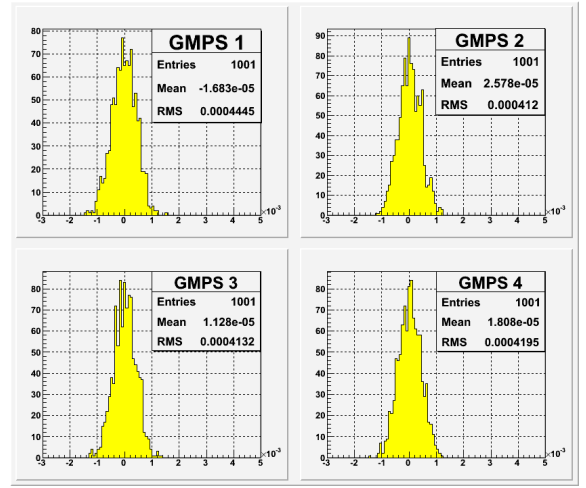}
    \caption{\label{fig:histo-noise-strength}
    Histogram of the amplitudes of noise 
    generated at each random noise node.}
    \end{center} 
  \end{figure}

\begin{figure}[hbt!]
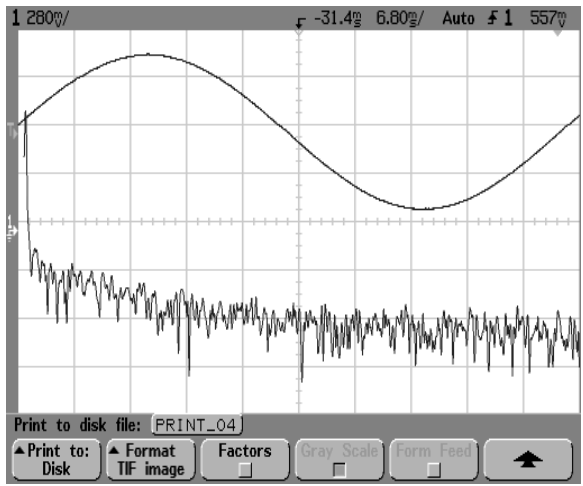

         \subfigure[ripple current]{\includegraphics[width=3.0in,height=2.5in]
            {Current_FFT_10kHz_1cycle.eps3}}
         \subfigure[Ornstein-Uhlenbeck noise]{\includegraphics[width=3.5in,height=3.in]
            {OU-Noise-5-ch4.eps3}} 
         \caption{\label{fig:fft-current}
                  (a)~FFT impulses and 15-Hz current waveform of one cycle; 
                  the horizontal scale is 167 Hz$/$div and  
                  the vertical scale is 20 dB$/$div.
                  (b)~The power spectral density of the Ornstein-Uhlenbeck noise 
                  is closely matched to that of measured ripple current} 
   \end{figure}

For stochastic noise models, the autocorrelation time $\tau_{ac}$
can be viewed as a memory span, or a measure of the dependence of 
the same stochastic values at two distinct times ($t$ and $t^{\prime}$).
In this subsection, the measured current fluctuations are parameterized
with the three stochastic parameters introduced in preceding 
sections~\ref{sec:process} and~\ref{sec:non-white-noise}:\\ 
\begin{compactenum}[(1)]\itemsep 0.2cm
   \item \textsfbf{time step ($\Delta t$)}: 
         The entire Booster magnet system is divided into four quadrants. 
         Each quadrant made up of a string of 24 magnets in series connection
         is driven by one GMPS. 
         Current fluctuations ($\Delta I/I$) from each GMPS 
         are transmitted to all magnets in each quadrant of the ring. 
         As such, all of the 24 magnets experience 
         the same amount of ripple current at an interval of the time step.
         Hence, the time step, or noise-sampling rate is chosen to be
         one revolution period ($T_{0}$ = 2.2 $\mu$s) 
         at injection energy of 400 MeV.
   \item \textsfbf{autocorrelation time, or correlation time} ($\tau_{ac}$):
         On the basis of direct current measurements from a main bus line,       
         the ripple currents are repeated above the base current, 
         or reference current at an interval of 1.5 $\sim$ 1.7 (ms)
         (see FIG.~\ref{fig:current-linear-waveform}). 
         Therefore, about the duration of 1.5 $\sim$ 1.7 (ms) 
         is chosen to be a proper autocorrelation time 
         for additional current fluctuations originated from each GMPS.
   \item \textsfbf{noise strength ($\mathcal{A}$)}:
         Based upon the amplitudes of ripple current 
         ($\Delta I/I$) on a linear ramping portion of 
         a sinusoidal current waveform 
         (cf. FIG.~\ref{fig:current-linear-waveform}),
         the rms value of fractional current fluctuation
         $\Delta I/I\Bigr\vert_{rms}$
         is on the order of $10^{-4}$. 
         For verification purpose, histograms of the O-U noise 
         generated from each of four noise nodes 
         that are symmetrically inserted around the Booster ring 
         are plotted. 
         As shown in FIG.~\ref{fig:histo-noise-strength},
         the rms values of histograms are on the same order
         as those of measured noise strengths.
\end{compactenum}  
%
The autocorrelation function of a signal, 
or the power spectra can be measured 
by means of FFT.
\begin{equation}
  S_{\xi}(\omega) 
  = \frac{1}{2\pi}\int_{-\infty}^{\infty}e^{-i\omega\tau}\mathcal{C}_{\xi}(\tau)\,d\tau
\end{equation}
\begin{equation}
  \mathcal{C}_{\xi}(\tau) 
  = \int_{-\infty}^{\infty}e^{i\omega\tau}S_{\xi}(\omega)\,d\omega
\end{equation}
According to the Wiener-Khinchine theorem~\cite{vankampen},
spectral density is the Fourier transform of 
the autocorrelation function~
$\mathcal{C}_{\xi}(\tau) = \langle~\xi(\tau)\xi(t~+~\tau)~\rangle$~
for stationary processes:
  \begin{equation}\label{eqn:wiener}
    S_{\xi}(\omega) 
  = 2\int_{0}^{\infty}\langle~\xi(\tau)\xi(t~+~\tau)~\rangle\cos(\omega\tau)\,d\tau,
  \end{equation}
with $S_{\xi}(\omega)$ being spectral density 
of a stochastic process $\xi$.
In FIG.~\ref{fig:fft-current}, FFT is performed 
with one-cycle range of time data from 15-Hz current.
For FIG~\ref{fig:fft-current}~(a), 
the horizontal scale is 167 Hz$/$div, 
and the vertical scale is 20 dB$/$div.
For FIG.~\ref{fig:fft-current}~(b),
the vertical scale is $10^{-1}$
to indicate power attenuation from 1.0.
The power spectral density of the O-U noise
is closely matched to that of the measured ripple current 
shown in FIG.~\ref{fig:current-linear-waveform}.
%
\subsection{\label{subsec:equivalent-model}
Equivalent-Circuit Model}
\begin{figure}[h!] 
     \begin{center}
        \includegraphics[scale=0.35]
        {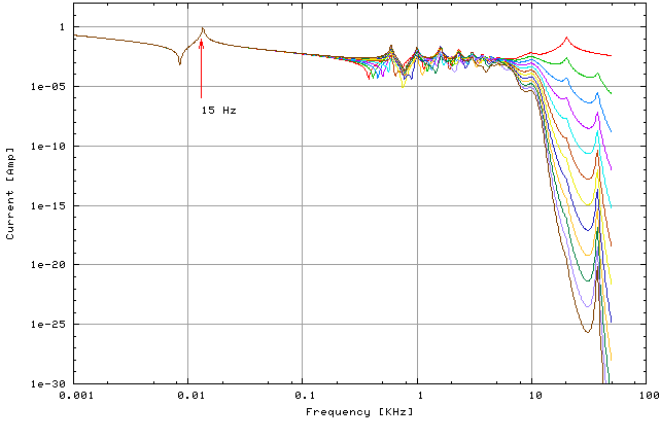}
        \caption{~\label{fig:spice}[SPICE simulation]:~Current vs. Frequency: 
                  current flowing through a string of 12 magnet cells 
                  driven by one GMPS. Progressing from top to bottom, 
                  the lines correspond to magnet cell [1] 
                  through magnet cell [12].}
     \end{center}
 \end{figure}

  To find out whether there are any offending resonances 
  floating around the Booster magnet system,
  acting as noise amplifiers, the equivalent circuit modeling 
  was also carried out. 
  The equivalent circuit 
  of one single LC-resonant cell is drawn 
  in Figure~\ref{fig:equiv-circuits}\cite{equiv}.
  Since a string of 24 magnets 
  in a quadrant of the Booster magnet system are connected in series, 
  they are treated as one transmission line. 
  We employed the B\sups{2} SPICE~\cite{spice} A/D Version 4, 
  which is one of many versions of commercial SPICE simulators. 
  The results of AC analysis of the equivalent circuits 
  are shown in FIG.~\ref{fig:spice}. 
  The current is peaked at 15 Hz and a cluster of minor peaks are 
  found in a few kHz range. 
  It is speculated that the offending resonances above 15 Hz 
  in higher frequency region could amplify the power supply noise,
  when the noise frequencies coincide with those of resonances. 
  The presence of the resonances could augment the formation of
  beam halo, eventually resulting in beam loss 
  during the injection cycle. 
  \begin{figure}[h!] 
     \centering
        \subfigure[Focusing Magnet]{\includegraphics[scale=0.16]{F-Magnet-bw.epsi}}
        \subfigure[Dedocusing Magnet]{\includegraphics[scale=0.18]{D-Magnet-bw.epsi}}
        \subfigure[Choke]{\includegraphics[scale=0.18]{Booster-Choke-bw.epsi}}
        \caption{\label{fig:equiv-circuits}
        Equivalent circuits of a focusing magnet, a defocusing magnet, 
        and a choke that comprise each magnet cell}
 \end{figure}
%
\section{\label{sec:tracking}TRACKING AND BEAM DIAGNOSIS}
%
%
\subsection{\label{subsec:sim-param}Simulation Parameters}
A comprehensive set of machine parameters for the Booster ring 
at injection energy is given in Table~\ref{tab:booster-param}.
Some parameters listed in Table~\ref{tab:booster-param} 
are derived from lattice parameters specified 
in the Booster design lattice (version 1.1).
\squeezetable
     \begin{table}[hbt!]
        \centering
        \caption{\label{tab:booster-param}
        Machine Parameters of Fermilab's Booster at Injection Energy}
        \vspace{\belowcaptionskip}
         \begin{tabular}{p{5cm} r}                  \\ \hline \\
         ring radius ($\langle R \rangle$)         & 75.47 (m)         \\
         ring circumference                        & 474.2 (m)         \\
         injection kinetic energy                  & 400 (MeV)         \\
         injection momentum                        & 954.263 (MeV$/$c) \\
         synchronous energy ($E_{s}$)              & 1.328 (GeV)       \\
         $\beta$ (Lorentz factor)                  & 0.7131            \\
         $\gamma$ (Lorentz factor)                 & 1.426             \\
         revolution period ($T_{0}$)               & 2.2 ($\mu$s)    \\
         revolution frequency ($f_{0}$)            & 454.5 (kHz)       \\
         no. of injection turns                    & 11                \\
         injection period                          & 24.2 ($\mu$s)   \\
         cycle time                                & 66.7 (ms)       \\
         $\gamma_{tr}$ (transition gamma)          & 5.4696            \\
         $\alpha_{1}$ (momentum compaction factor) & 0.0172            \\
         phase-slip factor ($\vert\eta\vert$)      & 0.458             \\ %
         $\varepsilon_{tr,~95,~n}$ (95~$\%$,~normalized)& 12.0 ($\pi$-mm-mrad) \\
         RF range                                  & 38.18 $\sim$ 52.83 (MHz)\\
         $\nu_{x0}$/$\nu_{y0}$ (bare tunes)        & 6.7 $/$ 6.8       \\
         betatron frequency ($f_{\beta,x}$, $f_{\beta,y}$) & 318.2 / 363.6 (kHz) \\
         $Q_{s}$ (synchrotron tune)                & 1.147 $\times 10^{-3}$\\
         $\Omega_{s}$ (synchrotron frequency)      & 3.28 (kHz)        \\
         $T_{s}$ (synchrotron period)              & 305 ($\mu$s)    \\
         $\sigma_{z}$~(rms bunch length)           & 1.0 (m) \\
         $\beta_{z}$ (longitudinal beta function)  & 3.0 $\times 10^{4}$ (m)   \\
         $\varepsilon_{l}$ (longitudinal emittance)& 0.25 (eV-s)       \\
         batch intensity                           & 5.04~$\times 10^{12}$ \\
         average beam current (at injection)       & 420 (mA)          \\
         effective beam radius                     & 0.0325 (m)        \\
         effective beam-pipe radius                & 0.0653 (m)        \\
         bunching factor ($B_{f}$)                 & $\sim$ 0.4        \\
         $\Delta\nu$ (tune shifts)                 & -~0.4             \\
         $\frac{\Delta P}{P_{0}}\bigr\vert_{max}$  & $\pm~0.15~\%$     \\
         $\sigma_{\delta}$                         & 3.0 $\times 10^{-4}$ \\
         $\beta_{x,max}~/~\beta_{y,max}$           & 33.7 $/$ 20.5 (m) \\
         $D_{x,max}~/~D_{y,max}$                   & 3.2 $/$ 0.0   (m) \\
         cell type                                 & FOFDOOD           \\
         cell length                               & 20.62 (m)         \\
         gradient magnets~/~cell                   & 4                 \\
         total gradient magnets                    & 96                \\
         $V_{rf,~inj}$ (RF voltage at injection)   & 205.0~(kV/Turn)   \\
         phase advance~/~cell                      & 96~(deg)          \\ 
         $\rho_{D}$ (defocusing bending radius)    & 48.034100  (m)    \\
         $\rho_{F}$ (focusing bending radius)      & 40.847086 (m)     \\
         \\ \hline
       \end{tabular}
   \end{table}
Salient ORBIT-FNAL simulation parameters 
including space-charge calculations are listed
in Table~\ref{tab:sim-param}.
\squeezetable
\begin{table}[hbt!]
\begin{center}
   \caption{\label{tab:sim-param}
   Salient Booster Simulation Parameters with ORBIT-FNAL}
    \begin{tabular}{p{5cm} r}
        \\ \hline                                 &           \\
        no. of injection turns                    & 11        \\
        no. of maximum macroparticles             & 330,000   \\
        harmonic no.                              & 84        \\         
        beam kinetic energy                       & 400.0 (MeV)  \\
        beam intensity~(per RF bucket)            & 6.0 $\times~10^{10}$\\
        transverse beam distribution              & bi-Gaussian  \\
        ring circumference                        & 474.2 (m)    \\
        $\beta_{x,~inj}$~$\vert$~$\beta_{y,~inj}$ & 6.274  / 19.312 (m) \\
        $\alpha_{x,~inj}$~$\vert$~$\alpha_{y,~inj}$ & -0.122 / 0.024    \\
        $D_{x,~0}~\vert~D_{y,~0}$                 & 2.581 $/$ 0.0 (m)\\
        $x_{0,~inj}$~$\vert$~$y_{0,~inj}$         & 0.0 / 0.0  (mm)  \\
        $E_{offset}$                              & 0.0 (GeV)        \\ 
        $\Delta E~/~E_{kinetic}$                  &  5.1 $\times~10^{-4}$ \\
        $\epsilon_{x,~rms,~inj}$~$\vert$~$\epsilon_{y,~rms,~inj}$ 
        & 1.76 / 1.76 ($\pi$-mm-mrad)                               \\
        $V_{rf}$ (RF voltage)                     & 205.0 (kV/Turn) \\
        $R_{wall}/R_{beam}$ (for geometric factor) & 2.0             \\
        longitudinal SC bin no.                   & 32              \\
        transverse SC bin no.                     & 64 x 64         \\ 
        smoothing parameter                       & $\thicksim$ 10\sups{-6}\\
        no. of total tracking turns               & 1,000           \\ 
                                                  &                 \\ \hline
     \end{tabular}
\end{center}
\footnotetext[1]{LSC stands for longitudinal space charge.}
\footnotetext[2]{TSC stands for transverse space charge.}
\end{table}
A round beam 
with axisymmetry is first injected into the Booster ring before tracking.
This ensures that we can solely investigate the noise effects 
under space charge alone. 
Optics functions ($\alpha(z)$, $\beta(z)$, $\gamma(z)$, $\delta(z)$)
are computed with the Booster design lattice using MAD (version 8.23)
prior to particle tracking.
According to the latest measurements and 
actual machine operation parameters,
a careful choice of the other simulation parameters
are made. 
\subsection{\label{subsec:sc-calc}
Parallelized Space-Charge Calculations}
\begin{figure}[hb!]
   \begin{center}
   \includegraphics[scale=0.35]{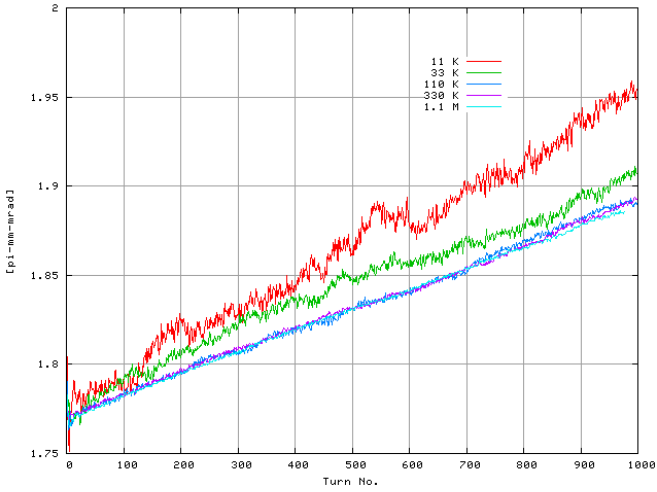}
   \caption{\label{fig:nmacros}
   Calculations of rms emittances with 
   a varying number of total \macros and 
   fixed space-charge bin numbers;
   progressing from top to bottom, 
   each trace corresponds with 
   11,000, 33,000, 110,000, 330,000, 
   and 1,100,000 \macros in total.}
   \end{center}
\end{figure}

A total of 330,000 macroparticles were 
 tracked for the full injection cycle of about 2.2 (ms). 
 A grid of 64 $\times$ 64 cells was used for 
 transverse space-charge calculations, 
 and 32 bins for longitudinal space-charge calculation  
 in the Particle-In-Cell (PIC) space charge model.
 During the course of tracking, 
 a total of 809 space-charge kicks were applied per revolution.
 This corresponds to about 2~kicks$/$m, or 58.6~cm$/$kick.
 In terms of betatron oscillations, about 121 kicks per 
 horizontal betatron oscillations, and about 119 kicks per
 vertical betatron oscillations, and about 17 kicks per magnet cell.
 Including both longitudinal and 
 transverse space-charge calculations,
 each parallelized calculation required about 8 hours 
 on forty-eight 2.0-GHz worker nodes.  
%
%
With space-charge bin numbers of 
$(64 \times 64)\times 32$ fixed,
rms emittances from tracking 
different numbers of macroparticles are calculated. 
As illustrated by FIG.~\ref{fig:nmacros},
when the total number of macroparticles 
amounts to above 330,000 after injection is complete, 
the time evolution of rms emittance converges 
with stability.
As such, considering practical computing time
and the number of macroparticles assigned to 
each space-charge bin, we determine that 
the number of 330,000 macroparticles is 
sufficiently large for accuracy.
Each macroparticle in the noise model
represents $\mathcal{O}$(10\sups{5}) 
real particles, or protons in the Booster.
%
Additional beam diagnostic calculations,
such as invariant action calculations, 
were implemented in \textit{parallel} mode. 
%
%
\subsection{\label{subsec:moments}Moments}
In the following subsections, 
we will present how beam diagnostic quantities
are defined and computed for the stochastic noise model.
we define moments which characterize
probability distributions of a beam, or macroparticles. 
Since it is necessary to consider beam centroids 
($\langle x \rangle$ and $\langle y \rangle$) in calculations, 
ORBIT-FNAL employs central moments:        
\begin{equation}\begin{aligned}\label{eqn:central-moments}
  \begin{cases}
     \Delta x_{r} &= x_{r} - \langle x_{r} \rangle\\
     \Delta y_{r} &= y_{r} - \langle y_{r} \rangle,
  \end{cases}
\end{aligned}\end{equation}
where $x_{r}(z)$ and $y_{r}(z)$ denote real-space coordinates. 
Because of vanishing central moment calculation,
beam centroids themselves ($\langle~x~\rangle$ and $\langle~y~\rangle$)
are used for the 1\sups{st} moment calculations.
It is assumed that the density profiles of an actual beam 
in transverse planes are {\em bi-Gaussian}. 
We first injected a herd of macroparticles of bi-Gaussian distribution.
Then, rms beam sizes ($\sigma_{x}$,~$\sigma_{y}$)
are calculated from the $2^{nd}$ moment calculation:
\begin{equation}\label{eqn:1st-moment}
\text{$1^{st}$ moments}
\begin{cases}
   \langle~x_{r}~\rangle\\
   \langle~y_{r}~\rangle
\end{cases}
\end{equation}
\begin{equation}\begin{aligned}\label{eqn:2nd-moment}
\text{$2^{nd}$ moments}
\begin{cases}
  \sigma^{2}_{x} &= \langle~(\Delta x_{r})^{2}~\rangle \\
  \sigma^{2}_{y} &= \langle~(\Delta y_{r})^{2}~\rangle 
\end{cases}
\end{aligned}\end{equation}
The rms beam sizes are important for space-charge study.
Starting with (bi-)~Gaussian charge distribution $\rho(r)$,
we can derive transverse space-charge force using Gauss' law 
and Amp\`{e}re's law. 
As given in Eqn.~(\ref{eqn:tsc-forces}),
the transverse rms beam sizes ($\sigma_{r}$)
determine the range of linear transverse space-charge forces 
($\mathcal{F}_{sc}(r)$):
\begin{equation}\label{eqn:tsc-forces}
  \begin{aligned}
    \begin{cases}  
  \rho(r) &= \frac{Ne}{2\pi\sigma_{r}^{2}}\cdot\exp\Bigl(-\frac{r^{2}}{2\sigma_{r}^{2}}\Bigr)\\
  \vec{\mathcal{F}}_{sc}(\vec{r}) &= \frac{Ne^{2}}{2\pi\varepsilon_{0}\gamma^{2}rL_{b}}
                         \Bigl(1 - exp\Bigl(-\frac{r^{2}}{2\sigma_{r}^{2}}\Bigr)\Bigr)\hat{r}\\
  r &= \sqrt{x^{2}+y^{2}}
    \end{cases}
  \end{aligned} 
\end{equation}
where $N$, $e$, $\varepsilon_{0}$, $\sigma_{r}$, 
$\gamma$, $r$, and $L_{b}$ 
are the number of particles per length, unit charge,
permittivity of the vacuum, rms beam size, 
Lorentz factor, radial distance, 
and bunch length, respectively. 
The transverse space-charge forces grow linearly with 
transverse displacements ($x$, or $y$),
and scale off with displacements larger than 2$\sigma_{r}$.
As the evolution of $2^{nd}$ moment calculations show
in presence of full space charge and power-supply noise,
the rms beam sizes grow steadily.
To understand the time evolution of 
rms beam sizes, or rms beam widths in transverse planes,
the 2\sups{nd} moments in real physical space are computed.
    \par The injection transverse coordinates employed in
    the ORBIT-FNAL
    are defined in physical space
    as a function of azimuthal coordinate $z$.
    Hence, the horizontal coordinates include 
    the effects of horizontal dispersion ($\mathcal{D}_{x0}(z)$).
    On the other hand, no dispersion effect is included 
    in the vertical coordinates because vertical dispersion 
    ($\mathcal{D}_{y0}(z)$) is set to zero in accordance with 
    the Booster design lattice. 
    Consequently, the following relations are implicitly reflected 
    in the macroparticle coordinates and the calculations of 
    transverse rms emittances:
\begin{equation}
\begin{aligned}\label{eqn:xreal-coordinates}
  \begin{cases}
  x_{r}(z) &= x_{\beta}(z) + D_{x}(z)\cdot\frac{\Delta p}{p_{0}}\\ 
  y_{r}(z) &= y_{\beta}(z)
  \end{cases}
\end{aligned}
\vspace{0.2in}
\end{equation}
%
In Eqn.~(\ref{eqn:xreal-coordinates}), 
$x_{\beta}(z)$ and $y_{\beta}(z)$ denote betatron coordinates,
and $D_{x0}(z)$ and $P_{0}$ denote injection horizontal dispersion 
and design momentum, respectively.
In a similar fashion, divergence angles are computed:
\begin{equation}\begin{aligned}\begin{cases}\label{eqn:yreal-coordinates}
  x^{\prime}_{r}(z) 
  &= x^{\prime}_{\beta}(z) 
   + D^{\prime}_{x}(z)\cdot\frac{\Delta p}{p_{0}}\\ 
  y^{\prime}_{r}(z) 
  &= y^{\prime}_{\beta}(z),
\end{cases}
\end{aligned}\end{equation}
where $\mathcal{D}^{\prime}_{x}(z)$ denotes 
the slope of horizontal dispersion.
%
\subsection{\label{subsec:rms-emit}RMS Emittances}
As the ORBIT-FNAL employs the $2^{nd}$-order central moments 
in the rms emittance calculations, we need to define 
additional quantities below to define rms emittances.  
  \begin{equation}\begin{aligned}\begin{cases}\label{eqn:displacement}
     \Delta \mathcal{P}_{x} &\equiv \mathcal{P}_{x} - \langle~\mathcal{P}_{x}~\rangle \\
     \Delta \mathcal{P}_{y} &\equiv \mathcal{P}_{y} - \langle~\mathcal{P}_{y}~\rangle 
  \end{cases}
  \end{aligned}\end{equation}
Once we define the $2^{nd}$-order central moments
of each coordinate in the 6-dimensional space,
we define column matrices 
$\mathcal{M}_{2,~x}$ and $\mathcal{M}_{2,~y}$.
%
\begin{equation}\begin{aligned}\label{eqn:col-xmatrix}
   &\mathcal{M}_{2,~x} =
   \begin{bmatrix}
     \Delta x_{\beta} \\
     \Delta x^{\prime}_{\beta}  
   \end{bmatrix}
   &\mathcal{M}_{2,~y} =
   \begin{bmatrix}
     \Delta y_{\beta} \\
     \Delta y^{\prime}_{\beta}  
   \end{bmatrix}
\end{aligned}\end{equation}
With the column matrices $\mathcal{M}_{2}$ above,
we can define 2 $\times$ 2 $\Sigma$-matrices 
in subspaces of trace space:
($x_{\beta},~x^{\prime}_{\beta}$)~and~($y_{\beta},~y^{\prime}_{\beta}$).
In each of $\Sigma$-matrices, 
the off-diagonal elements are associated with 
the correlation between position and angle, or 
energy and rf phase.
\begin{equation}\begin{aligned}\label{eqn:matrix}
  \mathlarger\Sigma(x_{\beta},~x^{\prime}_{\beta}) 
   &\equiv \langle\mathcal{M}_{2,~x_{\beta}}\mathcal{M}^{T}_{2,~x_{\beta}}\rangle\\
   &=
   \begin{bmatrix}
       \langle (\Delta x_{\beta})^{2}\rangle 
     & \langle \Delta x_{\beta} \Delta x^{\prime}_{\beta}\rangle \\
       \langle \Delta x^{\prime}_{\beta}\Delta x_{\beta}\rangle  
     & \langle {(\Delta x^{\prime}_{\beta}})^{2}\rangle
   \end{bmatrix}
\end{aligned}\end{equation}
in which $\mathcal{M}^{T}$ denotes a transpose matrix of $\mathcal{M}$.
Using the 2$\times$2 $\Sigma$-matrix, 
an {\em unnormalized rms emittance} can be defined as,
\begin{equation}\begin{aligned}\label{eqn:xemitsigma-1}
  \varepsilon_{x,~rms}
  & = \sqrt{\det\;\mathlarger\Sigma(x_{\beta},~x^{\prime}_{\beta})}\\
  & = \sqrt{\langle~(\Delta x_{\beta})^{2}~\rangle\langle~(\Delta x^{\prime}_{\beta})^{2}~\rangle 
    - \underbrace{\langle~\Delta x_{\beta}\Delta x^{\prime}_{\beta}~\rangle^{2}}_{\text{correlation term}}} 
\end{aligned}\end{equation}
Transverse rms emittances are defined 
in ($x_{\beta},~\mathcal{P}_{x}/P_{0}$) and 
($y_{\beta},~\mathcal{P}_{y}/P_{0}$) phase spaces, 
following the MAD~\cite{mad} convention:
\begin{equation}~\label{eqn:2xsigma}
  \mathlarger\Sigma(x_{\beta},~\mathcal{P}_{x})
   =
   \begin{bmatrix}
      \langle (\Delta x_{\beta})^{2}\rangle  & \langle\Delta x_{\beta} \Delta\mathcal{P}_{x} \rangle \\
      \langle \Delta \mathcal{P}_{x}\Delta x_{\beta}\rangle  & \langle(\Delta \mathcal{P}_{x})^{2} \rangle
   \end{bmatrix}
\end{equation}
\begin{equation}
\begin{aligned}\label{eqn:xemitsigma-2}
  \varepsilon_{x,~rms}
   &= \frac{1}{\mathcal{P}_{0}}\sqrt{\det\;
      \mathlarger\Sigma(x_{\beta},~\mathcal{P}_{x})}\\
   &= \frac{1}{\gamma\beta\,m_{0}c}
  \sqrt{\langle~(\Delta x_{\beta})^{2}~\rangle\langle~(\Delta\mathcal{P}_{x})^{2}~\rangle 
  - \langle~\Delta x_{\beta}\Delta\mathcal{P}_{x}~\rangle^{2}},\\
\end{aligned}
\end{equation}
in which the transverse momenta ($\mathcal{P}_{x}$ and $\mathcal{P}_{y}$) 
are normalized by the design momentum ($\mathcal{P}_{0}$). 
As stated earlier, if a unnormalized rms emittance 
is multiplied by the Lorentz factors ($\beta\gamma$), 
it transforms into a normalized rms emittance 
with no momentum dependence.
    The Eqns.~(\ref{eqn:col-xmatrix}) 
    through~(\ref{eqn:xemitsigma-2})  
    apply likewise to vertical and longitudinal planes.
\begin{equation}\begin{aligned}\begin{cases}\label{eqn:norm-emits}
  \varepsilon_{x,~n,~rms}
   &= (\beta\gamma)\varepsilon_{x,~rms}\\
   &= (\beta\gamma)
      \sqrt{\langle(\Delta x_{\beta})^{2}\rangle\cdot\langle(\Delta x^{\prime}_{\beta})^{2}\rangle
           -\langle~\Delta x_{\beta}\Delta x^{\prime}_{\beta}~\rangle^{2}}\\
   &= \frac{1}{m_{0}c}
      \sqrt{\langle(\Delta x_{\beta})^{2}\rangle\langle(\Delta\mathcal{P}_{x})^{2}\rangle
           -\langle~\Delta x_{\beta}\Delta\mathcal{P}_{x}~\rangle^{2}}\\
   \varepsilon_{y,~n,~rms}
   &= (\beta\gamma)\varepsilon_{y,~rms}\\
   &= (\beta\gamma)
      \sqrt{\langle(\Delta y_{\beta})^{2}\rangle\langle(\Delta y^{\prime}_{\beta})^{2}\rangle
           -\langle~\Delta y_{\beta}\Delta y^{\prime}_{\beta}~\rangle^{2}}\\
   &= \frac{1}{m_{0}c}
      \sqrt{\langle(\Delta y)^{2}\rangle\langle(\Delta\mathcal{P}_{y})^{2}\rangle
           -\langle~\Delta y\Delta\mathcal{P}_{y}~\rangle^{2}}
\end{cases}\end{aligned}\end{equation}
%
\section{\label{sec:gmps-noise}
        IMPACT OF GMPS CURRENT FLUCTUATIONS WITH SPACE CHARGE}
%
After inserting a total of four random-noise nodes,
provided with characteristic stochastic noise parameters,
into a Booster ring, 
macroparticles representing the Booster beams 
are tracked over 1,000 turns 
in the presence of full space charge\footnote{
~In the present context, 
full space charge is referred to as 
both transverse and longitudinal space charge,
or 3-D space charge}.

\begin{figure*}[ht!]
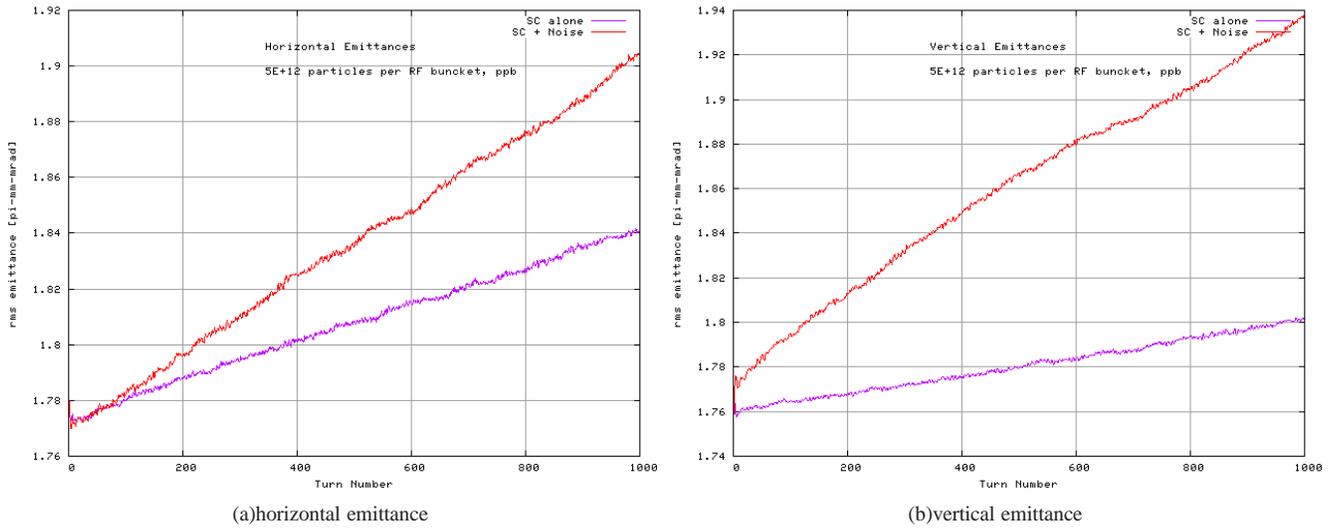

  \centering
     \subfigure[horizontal emittance]
        {\includegraphics[scale=0.35]{xemit-sc-brsc-5e12-mt.eps3}}
     \subfigure[vertical emittance]
        {\includegraphics[scale=0.35]{yemit-sc-brsc-5e12-mt.eps3}} 
     \caption{\label{fig:emit-noise-sc}
              transverse rms emittance growths starting from 
              the outset of injection through 1,000 tracking turns; 
              } 
\end{figure*}
\begin{figure*}[hb!]
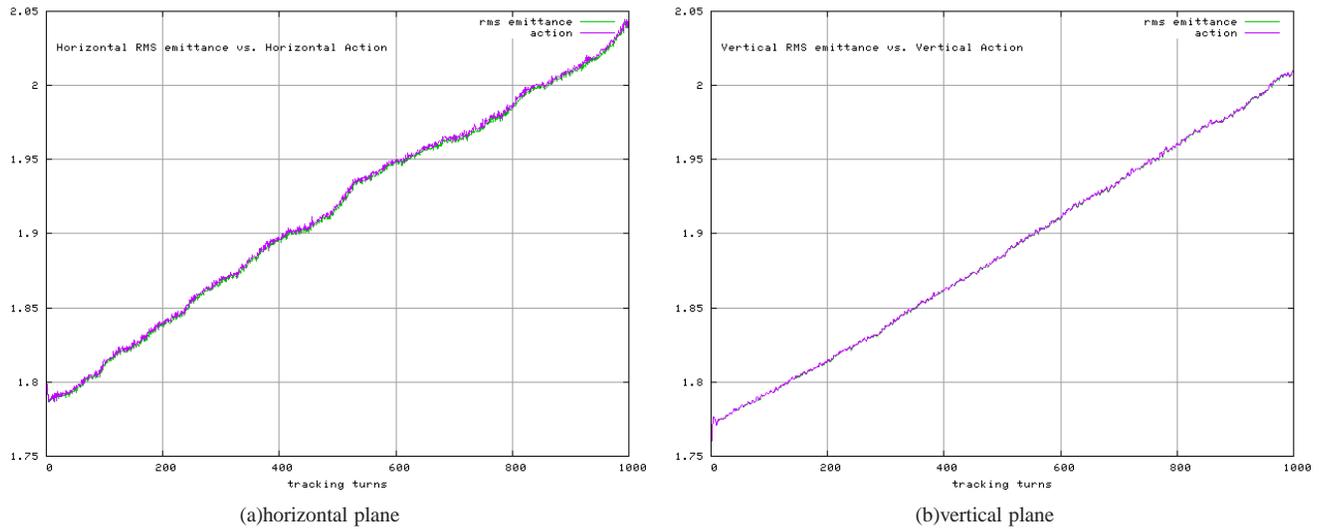

    \centering
       \subfigure[horizontal plane] 
         {\includegraphics[scale=0.35]{xactions-xemitt-noise.eps3}}
       \subfigure[vertical plane]
         {\includegraphics[scale=0.35]{yactions-yemitt-noise.eps3}}
       \caption{\label{fig:emit-action-noise}
        Time evolution of rms emittances in comparison with actions;
        (a)~horizontal rms emittance vs. horizontal action; 
        (b)~vertical rms emittance vs. vertical action}
\end{figure*}
%
\begin{figure*}[h!]
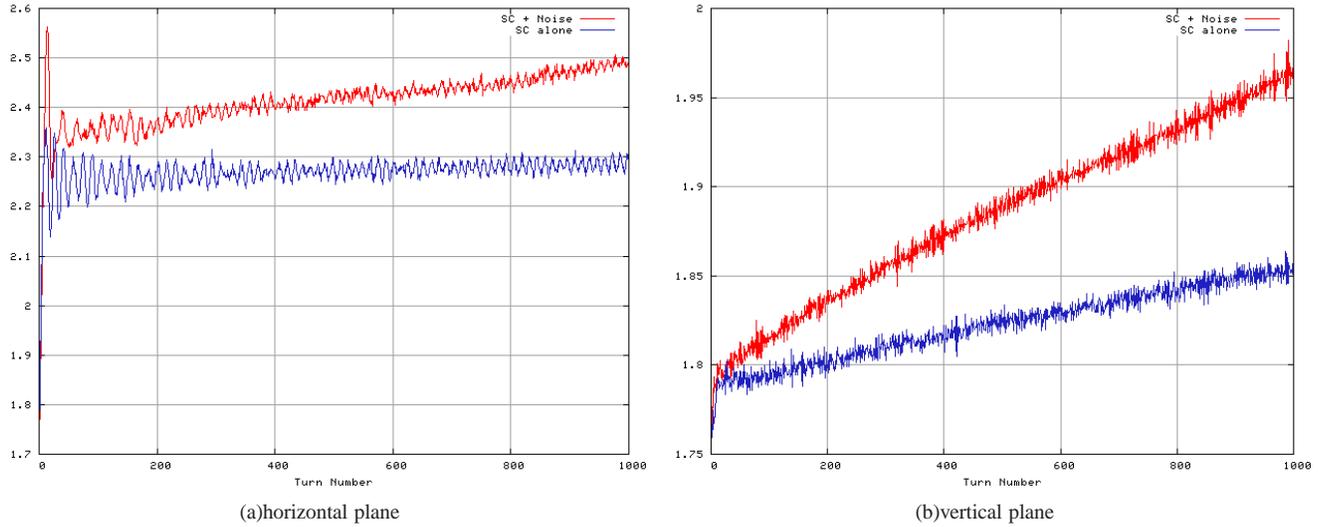

    \centering
       \subfigure[horizontal plane] 
         {\includegraphics[scale=0.35]{xmoments-sc-brsc.eps3}}
       \subfigure[vertical plane]
         {\includegraphics[scale=0.35]{ymoments-sc-brsc.eps3}}
       \caption{\label{fig:moments-brsc-sc}
        Time evolution of $2^{nd}$ moments in transverse planes; 
        full space charge alone (blue) and full space charge with noise (red)} 
\end{figure*}
\begin{figure*}[h!]
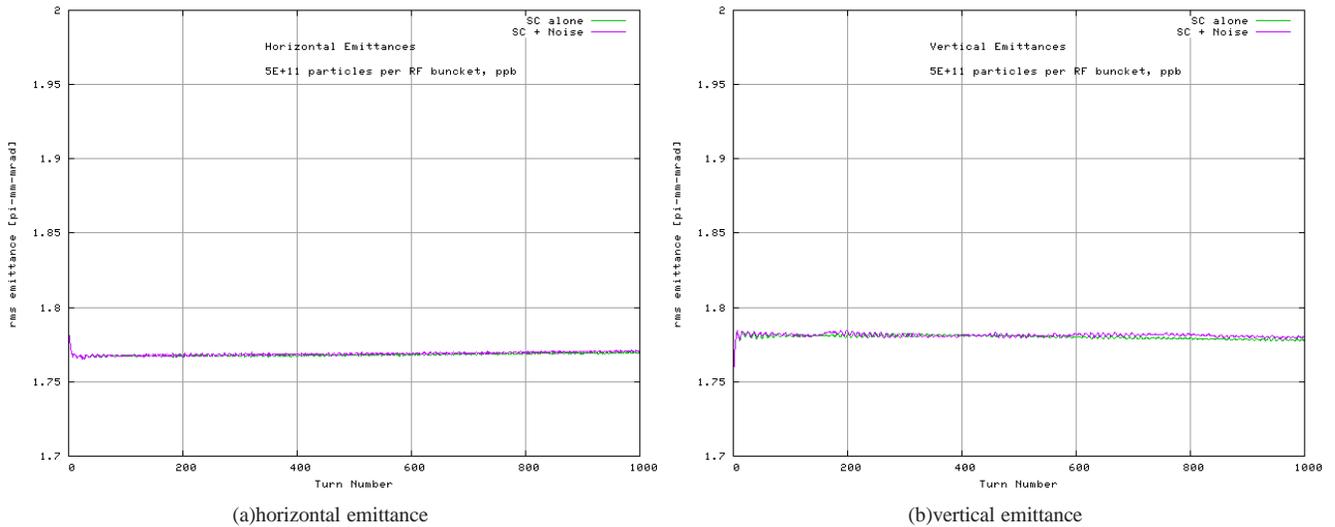
        
   \centering  
      \subfigure[horizontal emittance]
          {\includegraphics[scale=0.35]{xemit-sc-brsc-5e11.eps3}}
      \subfigure[vertical emittance]
          {\includegraphics[scale=0.35]{yemit-sc-brsc-5e11.eps3}}
     \caption{\label{fig:emit-lowint}
              transverse emittance growths; 
              the noise and space-charge effects in red and 
              the space-charge effects alone in blue. 
              The beam intensity is 6 $\times 10^{9}$ ppb, 
              and 5 $\times 10^{11}$ protons in total.}
\end{figure*}
\clearpage
\begin{figure}[htb!]
    \centering
       \includegraphics[scale=0.2]{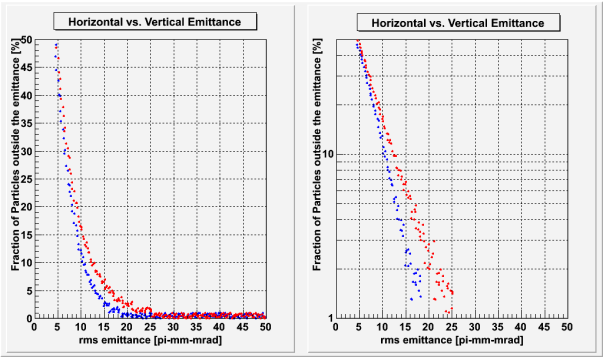}
       \caption{\label{fig:excl-action-histo}
                Fractional exclusion of macroparticles at a given average action. 
                The blue indicates at the $1^{st} turn$ 
                and the red indicates after 1,000 turns. 
                The vertical axis on the left plot is in linear scale,
                and the right is on logarithmic scale.}    
\end{figure}
 \begin{figure}[hbt!] 
    \centering
       \includegraphics[scale=0.2]{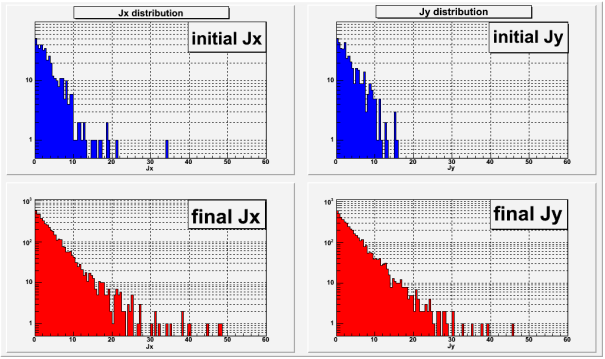}
       \caption{\label{fig:action-distro}The distribution of actions 
                ($\mathcal{J}_{x}$ and $\mathcal{J}_{y}$) 
                at the $1^{st}$ turn and after 1,000 turns. 
                O-U noise and 3-D space-charge effects are included.
                Action distribution at the $1^{st}$ turn is in blue, 
                and action distribution after 1,000 turns in red.}
 \end{figure}
%
%
As shown in FIG.~\ref{fig:emit-noise-sc},
the time evolution of transverse rms emittances
with the O-U noise\footnote{
~Hereafter, the O-U noise means 
the stochastic noise modeled on 
the GMPS noise measurements.}
coupled to the full space-charge effects (red) 
and with the space-charge effects alone (blue). 
The beam intensity per bucket is 6.0 $\times 10^{10}$ ppb, 
and the batch intensity is 5.0 $\times 10^{12}$ protons.
To estimate the emittance growth rate, 
the relative emittance growths
$\bigl(\large\frac{\Delta\varepsilon}
{\varepsilon_{0}}\normalsize\bigr)$~\footnote{
~$\varepsilon_{0}$ denotes initial emittance, and 
\begin{math}\Delta \varepsilon 
= \vert\varepsilon - \varepsilon_{0}\vert\end{math}}
are calculated starting 
from the last injection turns (the $11^{th}$ turn) 
through $1,000^{th}$ turn, 
prior to beam acceleration; 
this corresponds to the first 2 ms 
out of one cycle over 66.7 ms (15 Hz).
In the horizontal plane 
the relative emittance growth is about 7.5~$\%$, 
and in the vertical plane the growth is 9.3~$\%$. 
A total of 330,000 macroparticles, 
or 30,000 macroparticles per each injection turn
are simulated and tracked.
Upon including O-U noise representing the Booster GMPS noise
under space charge, the process of beam degradation develops, 
and a more noticeable halo formation is found. 
As a cross-check with the rms emittance calculations,
we also compute average actions at each tracking turn
including the noise and the full space-charge effects.
The rms emittances and average actions are overlaid
in FIG.~\ref{fig:emit-action-noise} for clear comparison.
The calculations of both rms emittances and actions 
manifest in such a good agreement 
that beam degradation is substantially enhanced 
due to \textit{synergistic mechanism} 
between GMPS-current fluctuations and space-charge effects.
Here, we use the term \textit{synergistic mechanism}
meaning that the total effects of GMPS noise and space charge 
are larger than the sum of individual effects.
The time evolution of rms beam sizes in both transverse planes
with space charge alone and with noise and space charge
are illustrated in FIG.~\ref{fig:moments-brsc-sc}. 
When the noise is included, the $2^{nd}$ moments, 
which are beam size squared, grow faster 
than in the case for space charge alone.
If we lower the Booster batch intensity 
by an order of one magnitude ($5 \times 10\sups{11}$)
from the present operational batch intensity
under the same conditions, 
the emittance growths induced by the GMPS noise 
and space-charge effects 
are not distinguishable from those of noise alone
in the absence of space charge 
as shown in FIG.~\ref{fig:emit-lowint}.
It should be noted 
that the space-charge effect is 
intensity dependent. 
Thus, if the beam intensity is lowered, 
so is the space-charge effects. 
This is a clear signature 
that only when the space-charge effects 
are substantial, so does the GMPS noise 
have a substantial impact on 
the Booster beam. 
In addition to the primary beam diagnostic calculations 
of the rms emittances and average actions,
we looked into the transverse couplings as well 
in the next subsection. 
%
\subsection{\label{subsec:couplings}Couplings}
\begin{figure*}[h!]
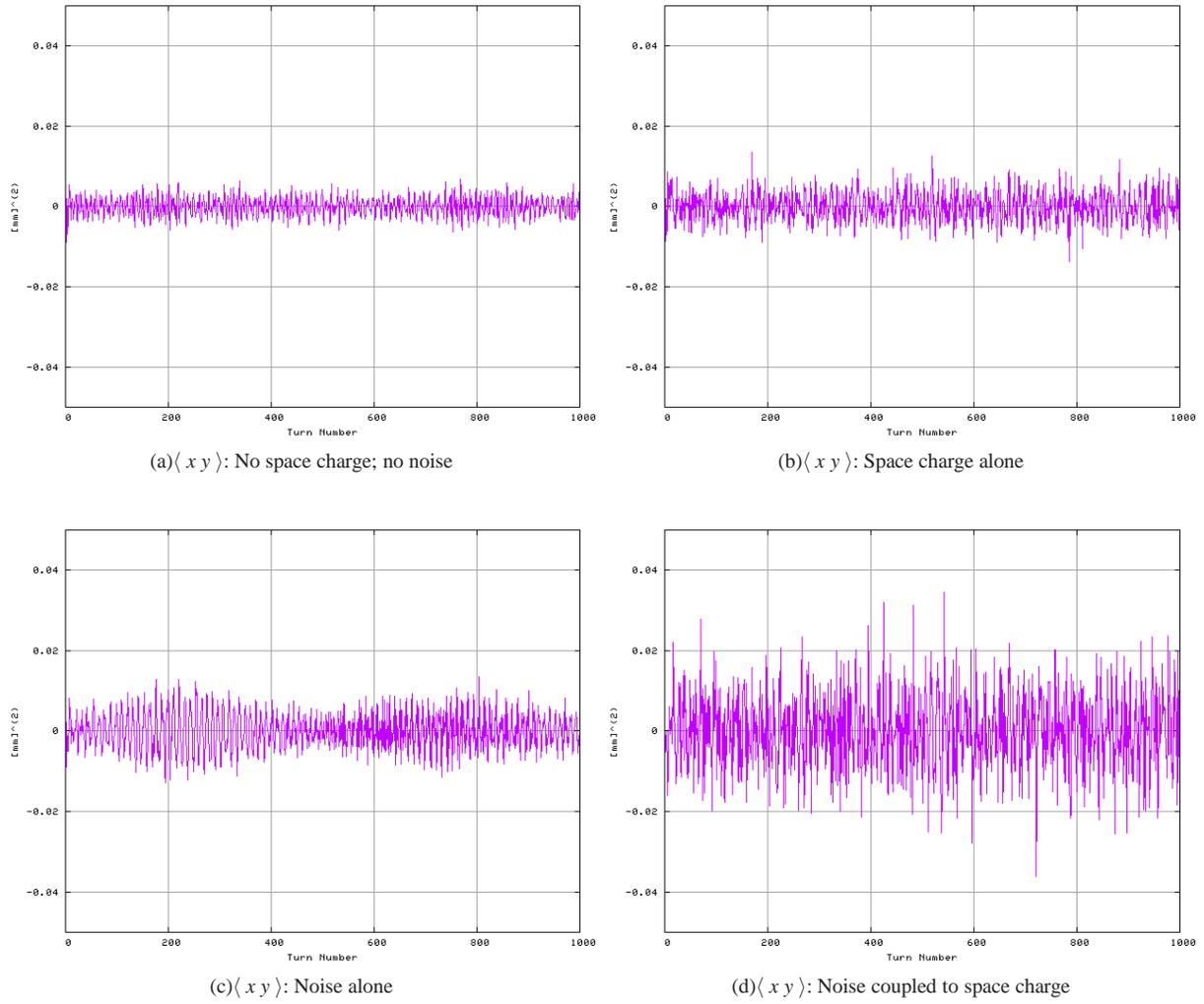
 
  \centering 
    \subfigure[$\langle~x~y~\rangle$:~No space charge; no noise]
    {\includegraphics[scale=0.32]{coupling-2d-nobrsc.eps3}}\vspace{0.08in}
    \subfigure[$\langle~x~y~\rangle$:~Space charge alone]
    {\includegraphics[scale=0.32]{coupling-2d-sc.eps3}}\vspace{0.08in}
    \subfigure[$\langle~x~y~\rangle$:~Noise alone]
    {\includegraphics[scale=0.32]{coupling-2d-br.eps3}}\vspace{0.08in}
    \subfigure[$\langle~x~y~\rangle$:~Noise coupled to space charge]
    {\includegraphics[scale=0.32]{coupling-2d-brsc.eps3}}
    \caption{\label{fig:coupling-2d}
    Transverse couplings in configuration space}
\end{figure*}

The computations of the $2^{nd}$-order cross moment 
$\langle~x~y~\rangle$ for each case are presented 
in FIG.~\ref{fig:coupling-2d}. 
A marginal amount of couplings are introduced 
due to the full space-charge effects 
(FIG.~\ref{fig:coupling-2d} (b)).
When the noise is included alone 
in the absence of the space charge,
couplings are somewhat noticeable 
(FIG.~\ref{fig:coupling-2d} (c)).
When the noise and the full space-charge effects are included, 
the transverse couplings are substantially amplified.
We therefore conclude that 
the noise impact on a beam 
is dependent upon the strengths of 
the space-charge defocusing forces 
in the Booster. 
What FIG.~\ref{fig:excl-action-histo}
illustrates is the percentage of 
macroparticles that reside outside of 
a given average action including
the O-U noise and space charge.
The blue markers indicate 
the fraction of excluded macroparticles
at a given emittance at the $1^{st}$ turn 
and the red markers at the end of tracking 
after 1,000 turns.
\cleardoublepage
FIG.~\ref{fig:action-distro} compares 
the distributions of transverse 
single-particle actions (\textit{$J_{x}$} and \textit{$J_{y}$}) 
at the outset of injection and at the end of 1,000 turns. 
It is evident that noise-induced beam degradation is enhanced 
as the time elapses.
\begin{figure}[ht!]
   \begin{center}
       \includegraphics[scale=0.25]{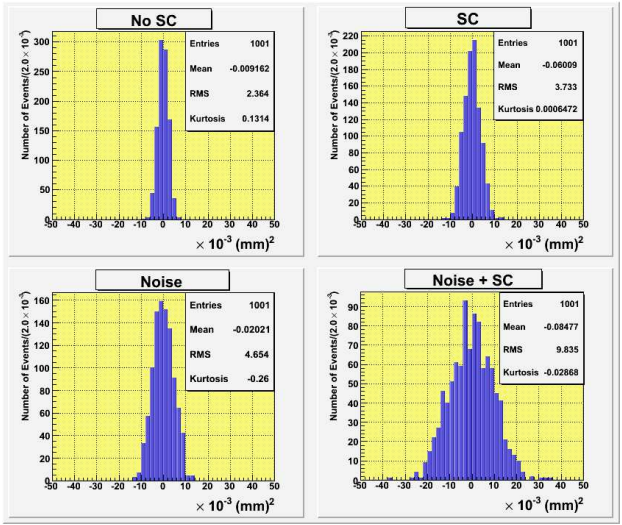}
       \caption{\label{fig:histo-cross-moment}
       The distributions of the magnitudes of 
       the $2^{nd}$-order cross moment ($\langle~xy~\rangle$)}
   \end{center}
\end{figure}
%
\subsection{\label{subsec:coupling-magnitude}Coupling Magnitude}
In an ideal system the normalized rms emittance remains constant.
However, nonlinear space-charge effect and couplings can induce 
degradation in beam quality. 
The increases of the normalized rms emittances indicate 
that nonlinear space-charge effect and couplings
induced by different machine imperfections are present in the Booster. 
One of the great advantages of the realistic accelerator simulation 
with macroparticle tracking is that
we can isolate an accelerator system condition 
to narrow down a specific cause of emittance growth under investigation.  
Therefore, in order to look into the transverse couplings, 
we additionally implemented in the ORBIT-FNAL 
new parallelized calculations of 4-dimensional transverse emittances 
($\varepsilon^{4}_{xy}$) and coupling magnitudes. 
From the determinant of 2$\times$2 \Large$\Sigma$\normalsize-matrix 
of beam distribution as given in Eqn.~(\ref{eqn:2xsigma}), 
a squared 2-dimensional rms emittance on the horizontal plane
can be calculated.
\begin{equation}\begin{aligned}\label{eqn:emit2sigma}
  \varepsilon^{2}_{x,~rms}
  &= \det\,
     \begin{vmatrix}
       \langle (\Delta x_{\beta})^{2}\rangle & \langle \Delta x_{\beta}\Delta\mathcal{P}_{x} \rangle \\
       \langle \Delta\mathcal{P}_{x}\Delta x_{\beta}\rangle  & \langle (\Delta\mathcal{P}_{x})^{2} \rangle
     \end{vmatrix}
\end{aligned}\end{equation}
Likewise, for the counterpart on the vertical plane.\\
For the computation of 4-dimensional rms emittances and couplings,
we first define 4-component column matrices 
($\mathcal{M}_{4,~xy}$, $\mathcal{M}_{4,~yz}$, and $\mathcal{M}_{4,~xz}$)
on two planes as in Eqn.~(\ref{eqn:col-matrix}).
By means of generating 4$\times$4 \Large$\Sigma$\normalsize-matrices
with the colum matrices as in Eqn.~(\ref{eqn:4d-sigma-matrix}), 
we can calculate the determinant of each 4$\times$4 matrix  
as described in Eqns.~(\ref{eqn:emit4xy}) through~(\ref{eqn:emit4xz}) 
that follow:
\begin{equation}\label{eqn:col-matrix}
  \mathcal{M}_{4,~xy} = 
  \begin{bmatrix}
    \Delta x_{\beta} \\
    \Delta\mathcal{P}_{x}\\
    \Delta y_{\beta} \\
    \Delta\mathcal{P}_{y}
  \end{bmatrix}
  \;
  \mathcal{M}_{4,~xz} = 
  \begin{bmatrix}
    \Delta x_{\beta} \\
    \Delta\mathcal{P}_{x}\\
    \delta \widetilde{\mathstrut E} \\
    \Delta \phi
  \end{bmatrix}
  \;
  \mathcal{M}_{4,~yz} = 
  \begin{bmatrix}
    \Delta y_{\beta} \\
    \Delta \mathcal{P}_{y}\\
    \delta \widetilde{\mathstrut E} \\
    \Delta \phi
  \end{bmatrix}
\end{equation}
 \begin{equation}\begin{aligned}\begin{cases}\label{eqn:4d-sigma-matrix}
     \mathlarger\Sigma_{xy} 
  &= \mathlarger\Sigma(x_{\beta},~x^{\prime}_{\beta},~y_{\beta},~y^{\prime}_{\beta})  
  = \langle\mathcal{M}_{xy}\mathcal{M}^{T}_{xy}\rangle \\
    \mathlarger\Sigma_{xz} 
  &= \mathlarger\Sigma(x_{\beta},~x^{\prime}_{\beta},~\delta \widetilde{\mathstrut E},~\phi)
  = \langle\mathcal{M}_{xz}\mathcal{M}^{T}_{xz}\rangle \\
    \mathlarger\Sigma_{yz} 
  &= \mathlarger\Sigma(y_{\beta},~y^{\prime}_{\beta},~\delta \widetilde{\mathstrut E},~\phi)
  = \langle\mathcal{M}_{yz}\mathcal{M}^{T}_{yz}\rangle,
 \end{cases}\end{aligned}\end{equation}
in which $\mathcal{M}^{T}$ denote a transpose matrix 
of $\mathcal{M}$.
\begin{equation}\label{eqn:emit4xy}
  \Large\varepsilon^{4}_{xy}\normalsize = 
    \det\,
    \begin{vmatrix}
       \langle(\Delta x_{\beta})^{2}\rangle & \langle\Delta x_{\beta}\Delta\mathcal{P}_{x}\rangle & 
       \langle\Delta x_{\beta}\Delta y_{\beta}\rangle & \langle\Delta x_{\beta}\Delta\mathcal{P}_{y}\rangle \\
       \langle\Delta\mathcal{P}_{x}\,\Delta x_{\beta}\rangle & \langle(\Delta\mathcal{P}_{x})^{2}\rangle & 
       \langle\Delta\mathcal{P}_{x}\Delta y_{\beta}\rangle & \langle\Delta\mathcal{P}_{x}\Delta\mathcal{P}_{y}\rangle \\
       \langle\Delta y_{\beta}\Delta x_{\beta}\rangle  & \langle \Delta y_{\beta}\Delta\mathcal{P}_{x}\rangle &
       \langle(\Delta y_{\beta})^{2}\rangle  & \langle\Delta y_{\beta}\Delta\mathcal{P}_{y}\rangle \\
       \langle\Delta\mathcal{P}_{y}\Delta x_{\beta}\rangle & \langle\Delta\mathcal{P}_{y}\Delta\mathcal{P}_{x}\rangle &
       \langle\Delta\mathcal{P}_{y}\Delta y_{\beta}\rangle & \langle(\Delta\mathcal{P}_{y})^{2}\rangle
    \end{vmatrix}
\end{equation}
In the same fashion, we can compute 
4-dimensional emittances and couplings on $x-z$ and $y-z$ planes as well.
\begin{equation}\label{eqn:emit4xz}
  \Large\varepsilon^{4}_{xz}\normalsize = 
    \det\,
    \begin{vmatrix}
       \langle(\Delta x_{\beta})^{2}\rangle    & \langle\Delta x_{\beta}\Delta\mathcal{P}_{x}\rangle & 
       \langle\Delta x_{\beta}\delta\widetilde{E} \rangle & \langle\Delta x_{\beta}\Delta\phi\rangle \\
       \langle\Delta\mathcal{P}_{x}\,\Delta x_{\beta}\rangle & \langle(\Delta\mathcal{P}_{x})^{2}\rangle & 
       \langle\Delta\mathcal{P}_{x}\delta\widetilde{E}\rangle & \langle\Delta\mathcal{P}_{x}\Delta\phi \rangle \\
       \langle\delta\widetilde{E}\,\Delta x_{\beta}\rangle & \langle\delta\widetilde{E}\Delta\mathcal{P}_{x}\rangle &
       \langle(\delta\widetilde{E})^{2}\rangle             & \langle\delta\widetilde{E}\Delta\phi\rangle \\
       \langle\Delta\phi\,\Delta x_{\beta}\rangle  & \langle\Delta\phi\Delta\mathcal{P}_{x}\rangle &
       \langle\Delta\phi\delta\widetilde{E}\rangle & \langle(\Delta\phi)^{2}\rangle
    \end{vmatrix}
\end{equation}
\begin{equation}\label{eqn:emit4yz}
  \Large\varepsilon^{4}_{yz}\normalsize = 
    \det\,
    \begin{vmatrix}
       \langle(\Delta y_{\beta})^{2}\rangle      & \langle\Delta y_{\beta}\Delta\mathcal{P}_{y}\rangle & 
       \langle\Delta y_{\beta}\delta\widetilde{E}\rangle   & \langle\Delta y_{\beta}\Delta\phi\rangle \\
       \langle\Delta\mathcal{P}_{y}\,\Delta y_{\beta}\rangle & \langle(\Delta\mathcal{P}_{y})^{2}\rangle & 
       \langle\Delta\mathcal{P}_{y}\delta\widetilde{E}\rangle& \langle\Delta\mathcal{P}_{y}\Delta\phi\rangle \\
       \langle\delta\widetilde{E}\,\Delta y_{\beta}\rangle & \langle\delta\widetilde{E}\Delta\mathcal{P}_{y}\rangle &
       \langle(\delta\widetilde{E})^{2}\rangle     & \langle\delta\widetilde{E}\Delta\phi\rangle \\
       \langle\Delta\phi\,\Delta y_{\beta}\rangle & \langle\Delta\phi\Delta\mathcal{P}_{y}\rangle &
       \langle\Delta\phi\delta\widetilde{E}\rangle& \langle(\Delta\phi)^{2}\rangle
    \end{vmatrix}
\end{equation}
Hence, {\em coupling magnitudes} between $x-y$, $y-z$, and $x-z$
can be calculated as follows:
\begin{equation}\begin{aligned}\begin{cases}\label{eqn:coupling}
  &\Delta\varepsilon^{4}_{xy} \\
  & = \Bigl\vert~\varepsilon^{4}_{xy}~-~\varepsilon^{2}_{x}\cdot\varepsilon^{2}_{y}~\Bigr\vert\\
  &  = \Bigl\vert\underbrace{\text{~$\mathlarger\sum$}\,\mathcal{C}_{xy}(~\langle\Delta x_{\beta}\Delta y_{\beta}\rangle, 
    \langle\Delta x_{\beta}\Delta\mathcal{P}_{y}\rangle,~\langle\Delta y_{\beta}\Delta\mathcal{P}_{x}\rangle,
   ~\langle\Delta\mathcal{P}_{x}\Delta\mathcal{P}_{y}\rangle~)}_{23~terms}~\Bigr\vert\\ \\
  &\Delta\varepsilon^{4}_{yz} \\
  &= \Bigl\vert~\varepsilon^{4}_{yz}~-~\varepsilon^{2}_{y}\cdot\varepsilon^{2}_{z}~\Bigr\vert\\
  &= \Bigl\vert\text{~$\mathlarger\sum$}\,
     \mathcal{C}_{yz}(~\langle\Delta y_{\beta}\,\delta\widetilde{\mathstrut E}\rangle, 
     \langle\Delta y_{\beta}\Delta\phi\rangle,~\langle~\delta\widetilde{\mathstrut E}\Delta\mathcal{P}_{y}\rangle,
    ~\langle\Delta\mathcal{P}_{y}\,\Delta\phi~\rangle~)~\Bigr\vert \\ \\
  & \Delta\varepsilon^{4}_{xz} \\
  & = \Bigl\vert~\varepsilon^{4}_{xz}~-~\varepsilon^{2}_{x}\cdot\varepsilon^{2}_{z}~\Bigr\vert\\
  & = \Bigl\vert\text{~$\mathlarger\sum$}\,\mathcal{C}_{yz}(~\langle\Delta x_{\beta}\delta\widetilde{\mathstrut E}\rangle, 
    \langle\Delta x_{\beta}\Delta\phi\rangle,~\langle~\delta \widetilde{\mathstrut E}\Delta\mathcal{P}_{x}\rangle,
   ~\langle\Delta\mathcal{P}_{x}\Delta\phi\rangle~)~\Bigr\vert
\end{cases}\vspace{0.2in}
\end{aligned}\end{equation}
%
where $\mathcal{C}_{xy}(\ldots)$ denotes coupling terms
as a function of 
$\langle\Delta x_{\beta}\Delta y_{\beta}\rangle$, 
$\langle\Delta x_{\beta}\Delta\mathcal{P}_{y}\rangle$,
$\langle\Delta y_{\beta}\Delta\mathcal{P}_{x}\rangle$, 
and $\langle\Delta\mathcal{P}_{x}\Delta\mathcal{P}_{y}\rangle$.
Hence, $\Delta\varepsilon^{4}_{xy}$ 
includes all possible combinations of couplings
not only between horizontal and vertical positions, 
but also between positions and divergence angles 
in transverse planes.
Furthermore, the 4-dimensional couplings 
can be extended to horizontal and longitudinal planes,
and to vertical and longitudinal planes.
What is illustrated by FIG.~\ref{fig:coupling-2d} is
the 2\sups{nd}-order cross moment of transverse positions 
($\langle\Delta x_{r}\Delta y_{r}\rangle$) 
to look into the transverse couplings
in the same vertical scale.
In the absence of space charge and GMPS noise, 
transverse coupling is not observed. 
In the presence of space charge, 
the magnitude of coupling is slightly increased
but still marginal.
However, with the GMPS noise alone 
in the absence of space charge, 
the coupling is more noticeable and 
some perturbation appear over 1,000 turns.
When the GMPS noise is coupled to 
the full space-charge effects,
the coupling is substantially amplified.
In FIG.~\ref{fig:histo-cross-moment}, 
the turn-by-turn calculations of the cross moments 
are presented in a form of histogram from which 
we extract statistics. 
The distributions are slightly dispersed
as each instability (either space charge, or GMPS noise) 
is individually included.
When the GMPS noise is applied to macroparticles
in the presence of space charge, 
the RMS value is larger than
that of the noise alone by about a factor of two. 
\begin{figure}[hb!]
       \includegraphics[scale=0.32]{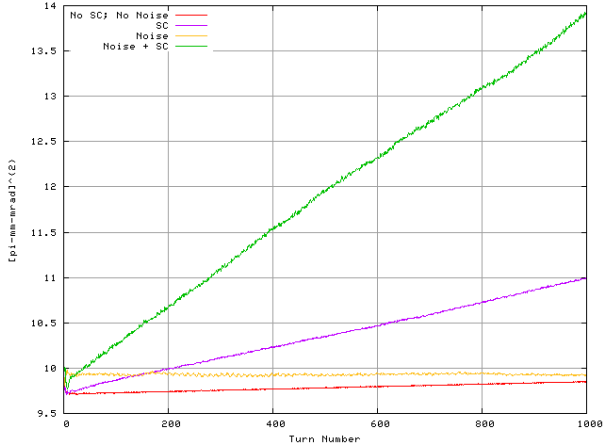}
       \caption{\label{fig:coupling}
       The time evolution of 4-dimensional coupling, 
       $\Delta\varepsilon^{4}_{xy}$}
\end{figure}
As derived in Eqn.~(\ref{eqn:coupling}), 
the coupling between horizontal and longitudinal planes
are continually growing when the GMPS noise and space charge
impinge on the Booster beam.  
In FIG.~\ref{fig:coupling}, 
progressing from bottom to top, 
each trace line corresponds with
each of the following cases:
(1) without space charge, nor GMPS noise,
(2) GMPS noise alone, (3) space charge alone, 
(4) GMPS noise in the presence of space charge.
In accordance with Eqn.~(\ref{eqn:coupling}),
the vertical axis is in units of ($\pi$-mm-mrad)\sups{2}.
It is evident from the FIG.~\ref{fig:coupling}
that transverse coupling is synergistically amplified
when the GMPS noise is coupled to full space-charge effects
in comparison with the other cases.
From the coupling calculations of 
$\Delta\varepsilon_{xy}$ and $\langle\Delta x_{r}\Delta y_{r}\rangle$,
we obtain consistent results;~\textit{the space charge
amplifies the impact of GMPS noise on the Booster beam}
%
\subsection{\label{subsec:halo}Halo Magnitudes}
\begin{figure}[hbt!]
    \includegraphics[scale=0.3]{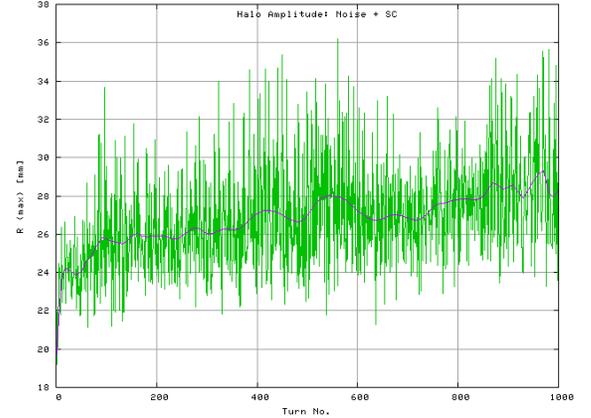}
    \caption{\label{fig:haloamp-bezier}Halo magnitudes ($R_{max}$): 
    noise in the presence of the space-charge effects; 
    the blue trace in the background indicates 
    smoothed curve with spline function.}
\end{figure}

\begin{figure}[hbt!]
    \includegraphics[scale=0.3]{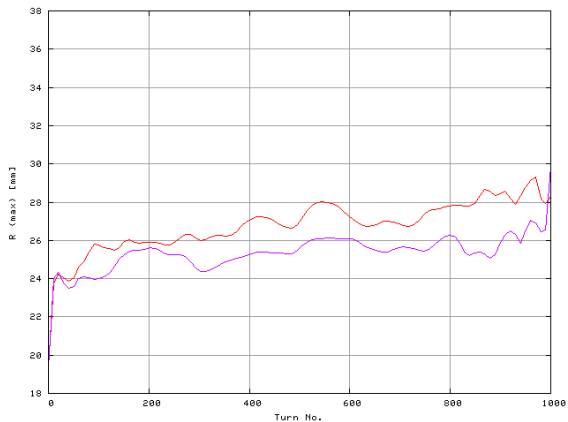}
    \caption{\label{fig:haloamp-smooth}
    Halo magnitudes ($R_{max}$): 
    noise along with space-charge effects (red) 
    vs. space-charge effects alone (blue)}
\end{figure}

The computation of maximum extent 
of macroparticle coordinates 
in a beam at each tracking turn is 
implemented in the Noise module.
The Eqn.~(\ref{eqn:haloamp-1}) includes 
only physical coordinates ($x$ and $y$) 
of a maximum-displaced macroparticle 
at the location of a random noise node~\cite{haloamp:clb}.
We refer it to as \textit{halo magnitude} ($R_{H,~2}$):
\begin{equation}\label{eqn:haloamp-1}
   \mathcal{R}_{H,~2}
   = \sqrt{x^{2} + y^{2}}\Bigr\vert_{Max}
\end{equation}
In Eqn.~(\ref{eqn:haloamp-2}),
~\textit{halo magnitude} in 4 dimension, 
($R_{H,~4}$), which includes horizontal and vertical
positions and angles of a maximum-displaced particle
is given:
\begin{equation}\label{eqn:haloamp-2}
  \mathcal{R}_{H,~4}\\
= \sqrt{(x/\sqrt{\beta_{x}})^{2} + (\sqrt{\beta_{x}}\cdot x^{\prime})^{2} +
        (y/\sqrt{\beta_{y}})^{2} + (\sqrt{\beta_{y}}\cdot y^{\prime})^{2}}
  \Bigr\vert_{Max},
\end{equation}  
where $\beta_{x}$ and $\beta_{y}$ are optics functions
at the location of a noise node.
Calculations of two types of halo magnitudes 
($\mathcal{R}_{H,~2}$ and $\mathcal{R}_{H,~4}$)
yield consistent results.
FIG.~\ref{fig:haloamp-bezier} illustrates 
the evolution of halo magnitudes in green 
and smoothed data in blue. 
Due to the large oscillatory behavior of 
the halo magnitudes, the data is smoothed.
The smoothed curve in FIG.~\ref{fig:haloamp-smooth}
~shows us with clarity a growing pattern of 
a maximum-displaced macroparticle 
from the physical center of a magnet aperture. 
%
%
\section{\label{sec:conclusion}DISCUSSIONS AND CLOSING REMARKS}
%
%
%
The model presented in this paper is
the first-ever measurement-based stochastic noise model
applied to an existing low-$\gamma$ accelerator lattice 
structure through stage approach.
%
Utilizing the state-of-the-art parallel computing technique
for better accuracy, we successfully incorporated and tracked 
a sufficiently large number of macroparticles
with FFT 3-D space-charge calculations
in a practical amount of computing time.
At first, the new noise module, which can generate a wide spectrum 
of stochastic noise ranging from white noise to colored noise,
was seemlessly integrated into the existing ORBIT-FNAL.
%
We then followed up with discovering 
the presence of a substantial amount 
of offending ripple current 
induced by common-mode voltage 
in the Booster power system.
However, on the other hand, 
the differential-mode voltage at each individual GMPS 
is well-smoothed with the aid of 
a 15-Hz low-pass filter
installed in each GMPS unit. 
Moreover, the root causes of 
the presence of common-mode voltage 
at each of four GMPS units were 
carefully diagnosed.
%
As a result of parameterization of
the Booster GMPS noise from 
the ripple-current measurements
with time step, autocorrelation time, and noise strength,  
FFT power-spectral densities between 
physical noise and modeled Ornstein-Uhlenbeck noise
are closely matched.
%
\par The foregoing results from particle tracking,
with the inclusion of power-supply noise as perturbation
and space charge as collective instability,
make evident that non-white noise 
originating from power supplies 
under the influence of space charge
leads synergistically to 
an enhancement of beam degradation 
phenomena---emittance growth, halo formation, 
and consequential beam loss---at the injection energy
of the Booster.   
%
As mentioned earlier, 
our investigations evidenced that 
the adverse effects of ripple current
are dependent upon the strength of space charge. 
Therefore, as a relevant side,
we can propose two approaches
to coping with the impact that 
ripple current has on charged-particle beams 
under the influence of space charge.
The first is to reduce inherent 
space charge forces themselves.
Over the past years, 
the efforts have been made to reduce 
the space-charge effects 
in the accelerator system at Fermilab. 
For instance, in 1993 
Fermilab's proton linac was upgraded from 
a beam kinetic energy of 200 MeV to 400 MeV
by adding more klystron tanks
in order to reduce the space-charge effects 
in the Booster. 
Besides, a dual RF system 
with a proper choice of RF parameters, 
allows us to further reduce space-charge effects
in high-intensity proton machines
by means of maneuvering charge distribution 
in longitudinal direction~\cite{tm-2368}.
Accordingly, attendant beam degradation phenomena
induced by fluctuating current and space charge 
can be suppressed.
The second approach is to devise 
instrumental techniques to cancel out 
common-mode-conducted EMI
originating from power supplies.
In particular, as demonstrated by 
the simulation of the equivalent-circuit model 
of the magnet system serving as an auxiliary model,
experimental measurements, or detection of
harmful high-frequency (HF) resonances residing 
in the magnet system (cf.~FIG.~\ref{fig:spice})
need to be pursued.
Once the presence of a cluster of 
parasitic HF resonances are confirmed, 
it is required that those resonances be damped out
to avoid the amplification of 
the adverse influence of power-supply noise 
on the Booster beam.
%
\par
Upon including more realistic and 
\textit{non-uniform} charge-density distribution~\cite{thesis:yoon},
the effects of space charge increases, 
so does the impact of current fluctuations
on the Booster beam accordingly. 
The modeling methodology presented in this paper
is expected to be well applicable to
other synchrotrons, or storage rings,
in which space-charge effects are of concern.
We therefore speculate that power-supply ripple current
can induce more prominent development of  
beam degradation process 
in storage rings of space-charge-dominated regime
over long period of time.
\vspace{0.1in}
%
\begin{acknowledgments}
  We are grateful to the University-Fermilab 
  Ph.D. Program Committee and 
  Prof. A.\ Bodek for their sponsorship, 
  and Dr. W.\ Chou of Fermilab 
  for his supervision~
  during the course of investigations 
  on the impact of fluctuating current on proton beams
  and modeling efforts.~
  ~The late Prof. C.\ L.\ Bohn of the Northern Illinois University 
  will be remembered for his steadfast encouragement 
  in efforts of modeling stochastic noise with realism.
  We also wish to thank Fermilab staff members of 
  Accelerator Division for helpful discussions and useful provisions,
  and of Theoretical Particle Physics Department and 
  of Computing Division for their arrangement 
  made to use local multi-CPU workstation clusters~\cite{lqcd}. 
  Special thanks should go to 
  the Accelerator Physics Group of SNS/ORNL 
  for their support in commissioning and upgrading
  the ORBIT-FNAL at the outset of the efforts.
  \par This work was supported by Fermi Research Alliances (FRA), LLC.,
  under the U.S. Department of Energy (DOE) contract No. DE-AC02-76-CH03000, 
  and by DOE Grant No. DE-FG02-91ER40685 to the University of Rochester.
\end{acknowledgments}
%
%
\cleardoublepage
%

%
\end{document}